\documentclass[reprint,nofootinbib,amsmath,amssymb,aps,pre,floatfix]{revtex4-2}

\usepackage{graphicx}
\usepackage{dcolumn}
\usepackage{bm}
\usepackage{amssymb}
\usepackage{hyperref}
\hypersetup{hidelinks}
\usepackage{braket}
\usepackage{cases}

\begin{document}

\title{Local Geometric and Transport Properties of Networks that are Generated from Hyperuniform Point Patterns}

\author{James V. Raj$^{1,*}$, Xiaohan Sun$^2$, Charles Emmett Maher$^{3,4}$, Katherine A. Newhall$^3$, and Mason A. Porter$^{2,5,6,7,8}$}

\affiliation{$^1$Department of Physics and Astronomy, University of California, Los Angeles, Los Angeles, CA, USA}\email[E-mail:] {james@physics.ucla.edu}
\affiliation{$^2$Department of Mathematics, University of California, Los Angeles, Los Angeles, CA, USA}
\affiliation{$^3$Department of Mathematics, University of North Carolina at Chapel Hill, Chapel Hill, NC, USA}
\affiliation{$^4$Department of Chemical and Biomolecular Engineering, Tandon School of Engineering, New York University, Brooklyn, NY, USA}
\affiliation{$^5$Department of Sociology, University of California, Los Angeles, Los Angeles, CA, USA}
\affiliation{$^6$Santa Fe Institute, Santa Fe, NM, USA}
\affiliation{$^7$Nordita, KTH Royal Institute of Technology, Stockholm University, and Uppsala University, Stockholm, Sweden}
\affiliation{$^8$Department of Mathematics, Stockholm University, Stockholm, Sweden}

\date{\today}

\begin{abstract}

Hyperuniformity, which is a type of order that is characterized by the suppression of long-range density fluctuations in comparison to the fluctuations in standard disordered systems, has emerged as a powerful concept
in many
natural and engineered phenomena. In this paper, we harness hyperuniform point patterns to generate
disordered, spatially-embedded networks that are distinct from both perfectly ordered lattices and uniformly random spatial
networks. We refer to this class of disordered networks as \emph{hyperuniform-point-pattern-induced (HuPPI) networks}, and we compare them to
counterpart \emph{Poisson-point-pattern-induced (PoPPI) networks} and networks that we generate from random-sequential-adsorption (RSA) packings (which have local order but are not hyperuniform).
 We examine HuPPI, PoPPI, and RSA networks that we generate using Gabriel, Delaunay, Delaunay-centroidal, and
Voronoi network-generation methods.
We show that HuPPI networks have smaller mean normalized total effective resistances (TERs)
than PoPPI networks in Gabriel, Delaunay-centroidal, and Voronoi networks and
have shorter mean random-walk mixing times than PoPPI networks in Gabriel, Delaunay, and Delaunay-centroidal networks.
For all four network-generation methods and for both HuPPI and PoPPI networks, we find that the Ollivier--Ricci curvature (ORC) of an edge is negatively correlated with its TER edge contribution (i.e., the increase in TER that results from removing that edge).
Additionally, PoPPI networks have wider distributions of TER edge contributions than their counterpart HuPPI networks.
The
edges in PoPPI networks with the largest TER edge contributions have negative ORCs and
act as critical bridges;  removing them
severely disrupts global transport.
The scarcity of such bridging edges in HuPPI networks helps explain
why they have smaller mean normalized TERs
than their PoPPI counterparts in Gabriel, Delaunay-centroidal, and Voronoi networks, including the counterintuitive observation that Delaunay-centroidal and Voronoi HuPPI networks have smaller mean normalized TERs despite having
more-negative mean ORCs.
Our comparison of HuPPI networks and PoPPI networks with RSA networks reveals some similarities between HuPPI and RSA networks in mean normalized TER and mean ORC, and it demonstrates that local order impacts the observed differences between HuPPI and PoPPI networks.
For all four network-generation methods, both HuPPI and PoPPI networks are less robust (i.e., the size of their largest connected component decays more rapidly)
when we remove edges in the order of their ORC values, starting with the most-negative ORCs, than when we remove edges uniformly at random.
In Gabriel networks, the HuPPI networks are more robust than their PoPPI counterparts for both edge-removal strategies.
Collectively, our results (1) illustrate how the local network geometry that ORC measures is related to transport and network connectivity and (2) demonstrate that the network-generation method
provides an additional design lever for spatial networks beyond the choice of their underlying point pattern.
\end{abstract}

\maketitle

\section{Introduction}\label{sec:Intro}

Networks arise in a wide variety of natural and engineered systems \cite{Newman2018}, and spatial embeddings and constraints play an essential role in many of these systems~\cite{barthelemy2022}. Disordered systems offer a much larger design space than ordered systems, enabling greater tunability of transport and mechanical properties, and recent work on spatially-embedded network metamaterials has begun to exploit this flexibility~\cite{ebra2026,wytock2026,Obrero2024,Liu2025,wang_local_topology_2026}.

From transportation infrastructure \cite{Barthelemy2024} and power grids~\cite{Cuadra2015} to
biological neural networks~\cite{Stiso2018} and protein-interaction networks \cite{Yang2015, albert2014}, spatial constraints play a fundamental role in shaping network structure and function.
The geometric arrangement of a network's nodes and the associated distances between them often constrain connectivity (e.g., by influencing the values of centrality measures~\cite{liu2020}), transport, and robustness~\cite{boguna_network_2021}.
Accordingly, there is considerable interest in understanding how different spatial point patterns, which range from uniformly random arrangements to highly correlated patterns, affect network properties after one uses such patterns to generate a network~\cite{barthelemy2022}. Indeed, the analysis of geometric networks that arise from point patterns is an important topic in the field of stochastic geometry~\cite{calka2009}.

One way to analyze the properties of geometric networks that one generates from point patterns is through geometric data analysis and the examination of discrete curvature~\cite{yadav2025}. Notions of network curvature have useful physical interpretations. For example, Devriendt
\emph{et al.}~\cite{dev2022,dev2024} introduced a resistance-based notion of network
curvature that links effective resistance, commute times, and mixing behavior in networks. Additionally, Robertson
\emph{et al.}~\cite{robertson2024} established a mathematical link between effective resistance and the $1$-Wasserstein distance in Ollivier--Ricci curvature (ORC)~\cite{ollivier2009,Ollivier2010} by demonstrating that they solve the same transport problem under different cost constraints. Motivated by this connection, we examine ORCs in spatially-embedded networks and explore the relationship between curvature properties and transport efficiency in such networks.
 We use ORC instead of Forman--Ricci curvature \cite{Sreejith_2016} (which is another popular type of network curvature) because the definition of ORC is based on optimal transport~\cite{Ollivier2010,Villani2009}, which aligns with our focus on network transport properties like total effective resistance (TER) and random-walk mixing time. The strong connection between effective resistance and optimal transport was also emphasized in a recent study of dissipation in slowly driven stochastic dynamics on networks~\cite{sawchuk2026}.

A key focus of our paper is to analyze the properties of \emph{hyperuniform-point-pattern-induced (HuPPI) networks}. A point pattern is \emph{hyperuniform} if its local density fluctuations are suppressed at large length scales in comparison to the fluctuations in a typical disordered system~\cite{torquato_local_2003,torquato_hyperuniform_2018}.
The importance of suppressed density fluctuations was recognized several decades ago, and it was analyzed in classical Coulomb systems (see, e.g., Refs.~\cite{Martin1980,Lebowitz1983,Martin1988,Jancovici1993,Levesque2000}) long before the term ``hyperuniform'' was coined by Torquato and Stillinger in Ref.~\cite{torquato_local_2003}.
In the last two decades, a large body of research
has demonstrated that hyperuniformity imparts desirable physical properties to both point patterns and two-phase media~\cite{torquato_hyperuniform_2018}.
(For a mathematical review of hyperuniformity, see Ref.~\cite{lach2025}.)
For example, Chen and Torquato~\cite{torquato_multifunctional_2018} demonstrated that certain types of disordered HuPPI networks can achieve almost optimal effective conductivity and elastic moduli, and they thereby highlighted transport advantages that hyperuniformity confers to networked two-phase media. Additionally, several researchers have demonstrated that disordered HuPPI dielectric network materials
can support isotropic photonic band gaps~\cite{Florescu2009,Man2013,Klatt2022,Siedentop2024}.
Very recently, Moody
\emph{et al.}~\cite{moody2026alloy} examined hyperuniformity in experiments with two-phase alloy microstructures.

Researchers are actively exploring how hyperuniformity manifests in spatially-embedded networks~\cite{maher_newhall_network_hyperuniform,Newby2025a,Newby2025b,wang2026, maher2026spectrum}. One approach is to use a point pattern, which can either be hyperuniform or nonhyperuniform,
to determine the locations of the nodes of a
network and to then assign edges between nodes using some geometric construction (e.g., a Delaunay or Voronoi tessellation).
Several researchers~\cite{Gabrielli2004,Newby2025a,Newby2025b} have illustrated that Delaunay and Voronoi
HuPPI networks
have narrower distributions of cell areas and smaller local density correlations than
their counterpart Poisson-point-pattern-induced (PoPPI) networks.
In the present paper, we investigate several properties of HuPPI networks, which we generate using Gabriel, Delaunay, Delaunay-centroidal, and
Voronoi network-generation methods.
HuPPI networks do not simply inherit the hyperuniformity of their underlying point patterns, but hyperuniformity
can nevertheless influence their properties.
For example, as demonstrated by Maher and Newhall~\cite{maher_newhall_network_hyperuniform}, the effective hyperuniformity of tessellation-based HuPPI networks
depends both on the network-generation method and on the disorder strength of the underlying point pattern.
Additionally, Vanoni
\emph{et al.}~\cite{Vanoni2025}
illustrated recently that the density-fluctuation suppression
in several hyperuniform systems occurs not only at large length scales but also
at intermediate and small length scales, and they suggested that such multiscale suppression can influence transport properties.
In our investigation, we observe differences in local geometric order between HuPPI and PoPPI networks.
We use the term ``local geometric order'' to signify similarity in
the structures of the local neighborhoods of different edges, rather than for positional order in the underlying point pattern.
We assess this similarity by calculating distributions of ORC values across edges.
Much is known about the transport properties of hyperuniform two-phase media \cite{Torquato2016, torquato_hyperuniform_2018, kim2021}, yet there is little understanding of the transport properties of HuPPI networks. Our aim is observational: we build HuPPI and counterpart PoPPI
networks, and we study whether and how the properties of the ensuing network families differ from each other.
We also compare HuPPI and PoPPI networks with networks that we generate from random-sequential-adsorption (RSA) packings, which have local order but are not hyperuniform.

We compute ORC \cite{Azarhooshang2020,Ollivier2010}, which is a discrete analogue of the Ricci curvature from
differential geometry~\cite{docarmo1992}, to determine locally how ``bottleneck-like" an edge is by comparing the optimal-transport cost that is required to
transform the probability distribution of the neighborhood of one of its attached nodes
into the probability distribution of the neighborhood of its other attached node.
When the neighborhoods of two adjacent
nodes share many common neighbors, the optimal-transport cost is low (i.e., the associated curvature is positive), whereas a lone bridge-like edge has a high transport cost (i.e., an associated negative curvature)~\cite{Azarhooshang2020, Ollivier2010,jost2014}.

We make three key sets of observations.
First, HuPPI networks have smaller mean normalized TERs~\cite{Ellens2011} than PoPPI networks in Gabriel, Delaunay-centroidal, and Voronoi networks and have shorter mean random-walk mixing times~\cite{montenegro2006} than PoPPI networks in Gabriel, Delaunay, and Delaunay-centroidal networks. Both the TER and the random-walk mixing time
are global measures of transport in networks.
Accordingly, our transport observations are consistent with previous findings about two-phase hyperuniform materials \cite{torquato_multifunctional_2018}, where the specific
network geometry and topology are the primary determinants of effective transport properties.
The different network-generation methods yield networks with distinct connectivity patterns.
Moreover, in several of our calculations,
the choice of network-generation method has a larger effect on the examined transport and curvature properties
than the choice between hyperuniform and Poisson point patterns.
Second, the RSA networks and HuPPI networks have some similarities in their mean normalized TERs and mean ORCs.
Because RSA packings have local order but are not hyperuniform, this comparison
demonstrates that local order impacts
the observed differences between HuPPI and PoPPI networks.
Third, mean ORC alone does not explain the observed differences in mean normalized TER.
Delaunay-centroidal and Voronoi HuPPI networks have smaller mean normalized TERs than their PoPPI counterparts even though they have more-negative mean ORCs.
In both HuPPI and PoPPI networks and for all four network-generation methods,
the ORC of an edge is negatively correlated with its TER edge contribution (i.e., the increase in TER that results from removing that edge).
Additionally, for all network-generation methods, the magnitude of this correlation is larger in HuPPI networks than in PoPPI networks.
The PoPPI networks have wider distributions of TER edge contributions than their HuPPI counterparts for all four network-generation methods.
The edges in the PoPPI networks with the largest TER edge contributions have negative ORCs and act as critical bridges, as removing them from a network severely disrupts global transport.
In Gabriel, Delaunay-centroidal, and Voronoi networks, the scarcity of such bridging edges in HuPPI networks helps explain their smaller mean normalized TERs.

To further explore the effects of removing edges in HuPPI and PoPPI networks, we examine the impact of two different edge-removal strategies on network connectivity.
We study such network robustness by scrutinizing
how the sizes of the networks'
largest connected components (LCCs)
decrease as we sequentially remove edges.
A network is more robust than another network if its relative LCC size (i.e., the fraction of its nodes that are in its LCC) is larger for the same fraction of removed edges.
For all four network-generation methods, HuPPI and PoPPI networks tend to
be less robust when we remove edges in the order of their ORC values (starting with the most-negative values) than when we remove edges uniformly at random.
In Gabriel networks, during most of the edge-removal process, the mean relative LCC sizes are larger for HuPPI networks than for their PoPPI counterparts for both
of our edge-removal strategies.

Our paper proceeds as follows. In Sec.~\ref{sec:Hyperuniformity}, we review the notion of hyperuniformity and discuss the standard classification of different types of hyperuniformity. In Sec.~\ref{sec:Methods}, we describe our procedures to generate HuPPI networks,
PoPPI networks, and RSA networks using Delaunay, Gabriel, Voronoi, and Delaunay-centroidal constructions. We also discuss how we compute ORCs, TERs, random-walk mixing times, and robustness and present our approach to assess network hyperuniformity.
In Sec.~\ref{sec:Global}, we compare the global transport properties of HuPPI networks,
PoPPI networks, and RSA networks and quantify the effective hyperuniformity of their associated two-phase network representations. We demonstrate that HuPPI networks have smaller mean normalized TERs than their counterpart PoPPI networks in Gabriel, Delaunay-centroidal, and Voronoi networks and shorter mean random-walk mixing times than their counterpart PoPPI networks in Gabriel, Delaunay, and Delaunay-centroidal networks. In Sec.~\ref{sec:OR}, we examine ORC statistics, including edge-level correlations between ORCs and TER edge contributions, to help interpret these global differences in terms of edge-level geometry.
In Sec.~\ref{sec:Robustness}, we evaluate network robustness for
HuPPI and PoPPI networks by removing edges both
 uniformly at random and in the order of their ORC values (starting with the most-negative values).
Finally, in Sec.~\ref{sec:Conc}, we conclude and discuss the implications of our results for the design of spatially-embedded networks.
The code to generate our point patterns and networks, analyze our networks, and generate our figures is available at \url{https://github.com/DMREF-networks/HuPPI-Network-Analysis} and is archived in the Zenodo repository~\cite{Raj2026HuPPICode}.
Our data are available in a searchable database of network configurations at \url{https://dmref-networks.github.io/disordered-configurations/}.

\section{Hyperuniformity}\label{sec:Hyperuniformity}

In this section, we define the notion of a hyperuniform point pattern in terms of number-variance scaling and present one method to generate such point patterns. Let $N(\Omega)$ be a random variable that gives
the number of points in a bounded spherical region $\Omega \subset \mathbb{R}^d$ of volume $|\Omega|$. We use $\rho$ to denote the \emph{number density}, which is the expected number of points per unit volume.
The expected number of points in $\Omega$ is $\langle N(\Omega)\rangle = \rho|\Omega|$ for a statistically homogeneous point pattern.
 The variance of the number of points is
\begin{equation}
	\sigma_N^2(\Omega) = \left\langle N(\Omega)^2\right\rangle - \langle N(\Omega)\rangle^2 \, .
\end{equation}
A point pattern is \emph{hyperuniform}~\cite{torquato_local_2003,torquato_hyperuniform_2018} (which was originally termed ``superhomogeneity"~\cite{gabrielli2002}) if the number variance of a region $\Omega_R$ satisfies
\begin{equation}
	\lim _{R \rightarrow \infty} \frac{\sigma_N^2\left(\Omega_R\right)}{\left|\Omega_R\right|} = 0 \, ,
\end{equation}
where $R$ is the radius (or any other linear measure of size) of a spherical observation window. That is, a point pattern is hyperuniform if the number variance grows more slowly than the window's volume. For example, if $\Omega_R$ is a ball of radius $R$, hyperuniformity requires the number variance to grow more slowly than $R^d$.

One can subdivide hyperuniform point patterns into three classes based on the leading-order scaling of $\sigma_N^2\left(\Omega_R\right)$ in the $R \rightarrow \infty$ (i.e., ``large-$R$") regime~\cite{torquato_local_2003,torquato_hyperuniform_2018}:
\begin{enumerate}
    \item Class I: $\sigma_N^2\left(\Omega_R\right) \sim R^{d - 1}$\,;
    \item Class II: $\sigma_N^2\left(\Omega_R\right) \sim R^{d - 1} \ln R$\,;
    \item Class III: $\sigma_N^2\left(\Omega_R\right) \sim R^{d - \alpha}$, where $0 < \alpha < 1$\,.
\end{enumerate}

In our investigation, we consider class-I hyperuniform point patterns that we generate using uniformly randomized lattices (URLs)~\cite{Gabrielli2004,klatt_cloaking_2020}, which we create as follows:
\begin{enumerate}
\item We start with a perfect two-dimensional (2D) square lattice in a square simulation box with side length $L$ and periodic boundary conditions. The spacing between adjacent lattice points is $1$. Such a lattice has strongly suppressed long-wavelength density fluctuations and thus is class-I hyperuniform.
Perfectly ordered lattices are trivially class-I hyperuniform.
\item We then shift each lattice point $\mathbf{r}_i = \left(x_i, y_i\right)$ by a vector $\boldsymbol{\delta}_i = \left(\delta_{i, x}, \delta_{i, y}\right)$, where we draw each entry of $\boldsymbol{\delta}_i$
from the uniform distribution on the interval $[-a/2, a/2]$. (It does not matter whether or not we include the end points $\pm a/2$ in the interval.)
\end{enumerate}

The perturbation strength $a$ controls the maximum displacement of the lattice points. Because we measure lengths in units of the lattice spacing, $a$ is a dimensionless parameter that encodes disorder strength. When $a = 0$, we have a perfect lattice with no translational disorder. As we increase $a$, the local translational disorder grows, yet the point pattern remains hyperuniform~\cite{klatt_cloaking_2020}.
Accordingly, we refer to $a$ as a ``disorder strength''.

\section{Methods}\label{sec:Methods}

In this section, we explain how we generate point patterns (see Sec.~\ref{subsec:PointGen}), construct networks from them (see Sec.~\ref{subsec:NetworkGen}), compute network measures (see Sec.~\ref{subsec:NetAnalysis}), and quantify effective network hyperuniformity from two-phase representations of networks (see Sec.~\ref{subsec:NetworkHU}).

\begin{figure*}[t]
  \centering
  \includegraphics[width=\linewidth]{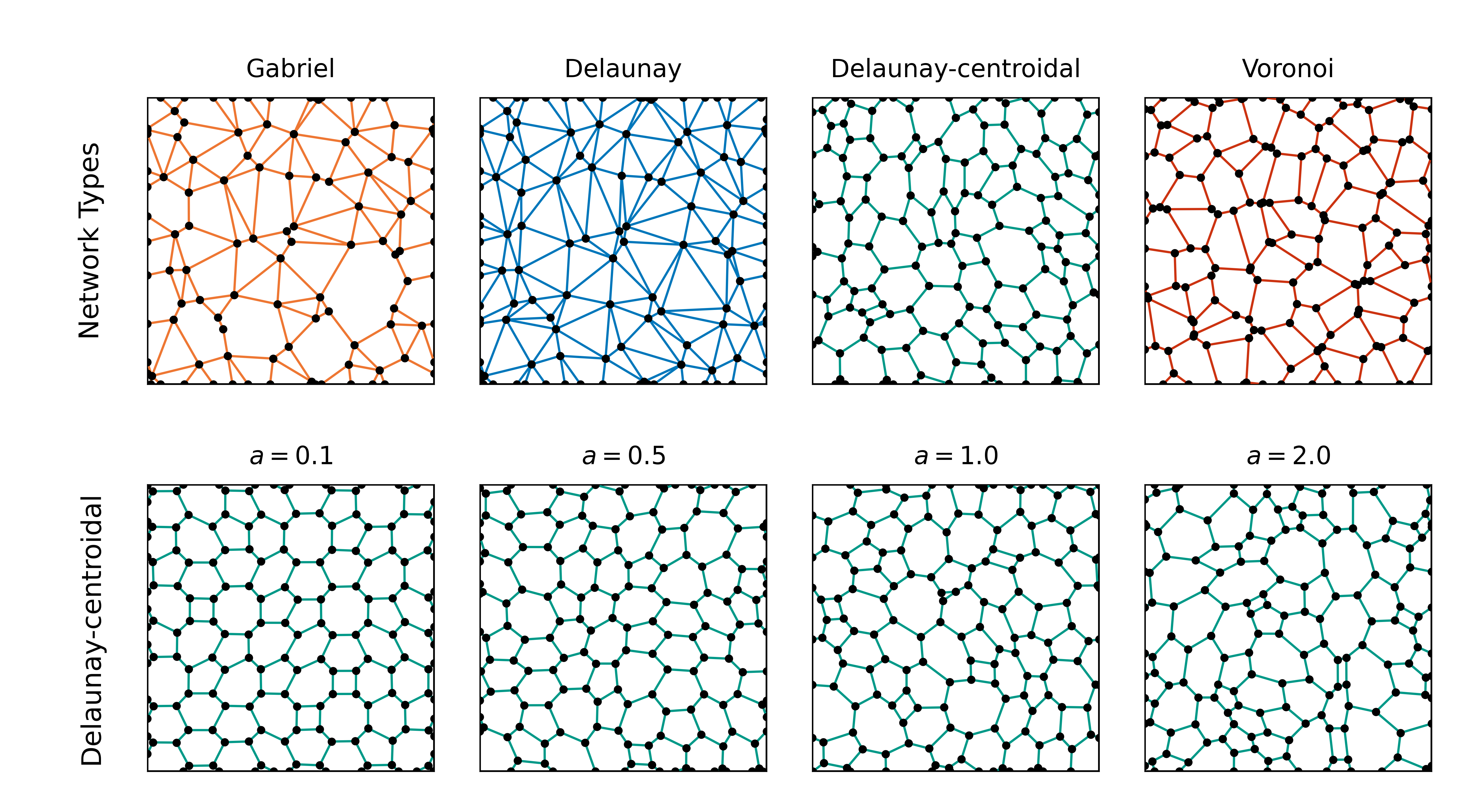}
  \caption{Visualizations of the types of networks that we construct from hyperuniform point patterns. In the top row, we show examples of Gabriel, Delaunay, Delaunay-centroidal, and Voronoi networks that we generate from a single point pattern with disorder strength $a = 1$. In the bottom row, we show Delaunay-centroidal networks that we generate using disorder strengths $a \in \{0.1, 0.5, 1, 2\}$, illustrating the transition from a near-lattice structure to a disordered structure. We generate and analyze networks with periodic boundary conditions, but we truncate edges at the boundaries in this figure to facilitate visualization.}
  \label{fig:network_visuals}
\end{figure*}

\subsection{Point-pattern generation}\label{subsec:PointGen}

We generate three distinct types of point patterns.
In our computations, we systematically
compare HuPPI networks (which we generate from hyperuniform point patterns), PoPPI networks (which we generate from Poisson point patterns), and networks that we generate from
random-sequential-adsorption (RSA) packings for several network-generation methods (see Sec. \ref{subsec:NetworkGen}).
As a convention, we use the term ``system size'' for the number of points ($N_{\mathrm p}$) in a point pattern and we use the term ``network size'' for the number of nodes ($n$) in a network.

We generate ensembles of
class-I hyperuniform point patterns
for disorder strengths
$a \in \{0.1, 0.2, \ldots, 2\}$ inside a square periodic domain $[0,L) \times [0,L) \subset \mathbb{R}^2$. To have a
counterpart nonhyperuniform point pattern for each system size, we generate
Poisson point patterns by uniformly random sampling points in the same 2D periodic domain with the same number density as the corresponding hyperuniform patterns.
The Poisson point patterns are not hyperuniform, as their number variances scale linearly with the window volume (i.e., $\sigma_N^2(\Omega_R) \sim |\Omega_R|$ as $R \rightarrow \infty$), in contrast to the suppressed fluctuations in hyperuniform systems.

To further clarify the relative roles of local geometry and hyperuniformity, we also generate point patterns using RSA
packings of
monodisperse disks.
For a recent review of RSA, see Ref.~\cite{Kubala2022RSA}. In RSA, one samples trial disk centers uniformly at random and accepts a trial disk if and only if it does not overlap any existing disk;
otherwise, one rejects the trial disk and samples a new trial location for a disk. We use sets of centers of accepted disks as point patterns.
We construct RSA packings with packing fraction $\phi = 0.53$, which is close to the 2D saturation limit $\phi_{\mathrm{sat}} \approx 0.547$ at which it is impossible to insert an additional disk
without creating an overlap between disks~\cite{Kubala2022RSA,ZhangTorquato2013,torquato2026}.
In our comparisons, we ensure that each
RSA packing has exactly
$N_{\mathrm p}$ disks, which matches the number of points in the HuPPI and PoPPI patterns.
Near-saturated RSA packings are useful for our comparisons because their nonoverlap constraint induces local order through correlations between nearby disk centers, but RSA packings are
not hyperuniform~\cite{TorquatoUcheStillinger2006}.
Therefore, they provide an informative comparison with the Poisson point patterns and the class-I hyperuniform point patterns that we generate from URLs.

\subsection{Network generation}\label{subsec:NetworkGen}

There are numerous models of random tessellations~\cite{Redenbach2025}.
In our study, we construct four types of planar networks (and associated tessellations) from the 2D
hyperuniform, Poisson, and RSA point patterns that we described in Sec.~\ref{subsec:PointGen}.
We show examples of the HuPPI
networks in Fig.~\ref{fig:network_visuals}.
For each point pattern $P = \{\mathbf{p}_1, \mathbf{p}_2, \ldots, \mathbf{p}_{{N_{\mathrm p}}}\} \subset \mathbb{R}^2$, we generate four families of networks.
\begin{enumerate}
\item\textbf{Delaunay tessellations~\cite{barthelemy2022}}: We consider each set $\{\mathbf{p}_i, \mathbf{p}_j, \mathbf{p}_k\} \subset P$ of three points, and we form edges between each pair of points in the set to form a triangle
$[\mathbf{p}_i, \mathbf{p}_j, \mathbf{p}_k]$ if and only if the circumscribed circle of the set includes no other points from $P$ in its interior. Delaunay tessellations have been employed in the design of disordered
mechanical
metamaterials with programmable static and dynamic responses~\cite{Chen2024MNN}.
\item\textbf{Gabriel networks}~\cite{barthelemy2022}: We connect two points $\mathbf{p}_i$ and $\mathbf{p}_j$ if and only if the closed disk whose diameter is the line segment $\overline{\mathbf{p}_i\mathbf{p}_j}$ includes no other points from $P$. We refer to this disk as a \emph{diametral disk} and to the requirement that there are no other points from $P$ within it as an ``empty-disk requirement". A Gabriel network is a subgraph of a Delaunay tessellation.

\item\textbf{Voronoi tessellations}~\cite{barthelemy2022}: In a periodic square domain $[0,L) \times [0,L) \subset \mathbb{R}^2$, the Voronoi cell $V_i$ of each point $\mathbf{p}_i \in P$ is the set of all points $\mathbf{x} \in [0,L) \times [0,L)$ that are closer to $\mathbf{p}_i$ than to any other point in $P$. That is, $V_i = \{\mathbf{x} \in [0,L) \times [0,L) \mid \|\mathbf{x} - \mathbf{p}_i\| \leq \|\mathbf{x} - \mathbf{p}_j\| \text{ for all } j \neq i\}$. Each node is attached (i.e., ``incident") to exactly three edges, so the resulting network
is 3-regular. The three edges that meet at a
node intersect only at that node, so incident edges never overlap along their interiors. We use the term ``tree-like" to describe this
local geometry, which is responsible for the predominantly negative ORCs that we observe in Voronoi networks (see Sec.~\ref{sec:OR}).
Voronoi tessellations of randomly distributed
points have been employed in the design of strut networks in mechanical metamaterials to achieve
tunable energy absorption~\cite{Cronau2025}.
\item\textbf{Delaunay-centroidal tessellations}\footnote{Centroidal Voronoi tessellations are a related but distinct tessellation type
in which each Voronoi-cell generator coincides with the centroid of its cell~\cite{du1999}. By contrast, Delaunay-centroidal networks use the centroids of Delaunay triangles as nodes.}\!\!~\cite{torquato_multifunctional_2018}: We first compute the Delaunay tessellation of a point pattern, and we then compute the centroid
$\displaystyle \mathbf{c}_{\,i_1, i_2, i_{3}} = \frac{1}{\,3\,}\sum_{m = 1}^{3}\mathbf{p}_{\,i_m}$ of each of its triangles $[\mathbf{p}_{i_1},\mathbf{p}_{i_2},\mathbf{p}_{i_3}]$.
We connect two such centroids if and only if their corresponding triangles share a common edge in the Delaunay tessellation. Each triangle has exactly three edges, so each centroid is adjacent to precisely three neighbors, yielding a
3-regular network
in which the edges that are incident to a
node intersect only at that node. This tree-like structure also causes most edges to have negative ORCs (see Sec.~\ref{sec:OR}).
Delaunay-centroidal tessellations have been employed in studies of photonic band gaps in disordered HuPPI dielectric network materials~\cite{Florescu2009,Man2013,Klatt2022}.

\end{enumerate}

These four generation methods yield networks with distinct
features even when they have the same underlying point pattern. To compare their transport properties on equal footing, we prescribe a common weighting scheme for their edges. We treat all networks as undirected, and we weight their edges by assigning edge conductances that depend solely on edge lengths.

For the Delaunay-centroidal and Voronoi network-generation methods,
it is crucial to distinguish between the nodes of the final network and the points of the initial point pattern.
 In a Delaunay-centroidal network, the nodes are the centroids of the Delaunay triangles. In a Voronoi network, the nodes correspond to the
circumcenters (i.e., the centers of the circumscribed circles) of the Delaunay triangles.
In both types of networks, two nodes are adjacent if and only if their associated Delaunay triangles share an edge. Therefore,
Delaunay-centroidal and Voronoi networks that we generate from the same point pattern have the same nodes and edges; they
differ only in the locations of their nodes (and hence in the lengths and weights of their edges).
For both Delaunay-centroidal networks and Voronoi networks, the number $n$ of nodes (i.e., the network size)
is directly related to the number $N_{\mathrm p}$ of points in the underlying point pattern (i.e., the system size) by the Euler-characteristic formula for a torus \cite{doCarmo1976}.
In a generic configuration, this formula yields $n = 2N_{\mathrm p}$. Therefore, Delaunay-centroidal and Voronoi networks have twice as many nodes as the number of
points in the underlying point pattern. By contrast, $n = N_{\mathrm p}$ for the Delaunay and Gabriel network-generation methods
because the points themselves are the nodes.
For generic periodic tessellations, Delaunay, Delaunay-centroidal, and Voronoi networks have $3N_{\mathrm p}$ edges. By contrast, Gabriel networks can have different numbers of edges for the same system size
$N_{\mathrm p}$ because satisfying the empty-disk requirement can remove different numbers of Delaunay edges. All $N_{\mathrm p}$ nodes remain after removing these edges.

Given a network (i.e., graph) $G = (V,E)$ that is embedded in a square domain with periodic boundaries, a ``cell" is a connected open set $C \subset \mathbb{R}^2$ whose boundary $\partial C$ is a simple closed polygon with edges from the set $E$. Distinct cells have disjoint interiors and the union of their closures covers a
 domain, so the cells are precisely the 2D faces between the nodes and edges
of $G$. In a Delaunay triangulation, each triangular face is a cell. In a Gabriel network, the cells are the
faces (which need not be triangles) of
the network.
Each Voronoi region is a cell. The cells of a Delaunay-centroidal network are the
faces that one obtains
by connecting the centroids of a Delaunay triangulation, with each cell of a
Delaunay-centroidal network centered on a node of its associated
Delaunay triangulation.

\subsubsection{Edge weights and conductance}

We assign a conductance to each edge $(i, j) \in E$ of a network
to give it a weight $w_{ij} = 1/\ell_{ij}$, where $\ell_{ij}$ is the Euclidean length of the edge. With this weighting, shorter edges have larger conductances, which is physically sensible for transport processes.

\subsubsection{Periodic boundary conditions}

We seek to study large-scale density fluctuations and avoid finite-size effects at the boundaries, so we impose periodic boundary conditions for all point patterns prior to generating networks from them.  Concretely, we identify (1) the left and right boundaries of the square domain and (2) the top and bottom boundaries of the square domain, effectively ``wrapping'' the system to form a torus. Consequently, the edges of the Gabriel, Delaunay, Delaunay-centroidal, and Voronoi networks can connect points that lie near opposite sides of our computational box.

\subsubsection{Network ensembles and sampling}

For each set of parameters (i.e., system size, disorder strength $a$, and point-pattern type), we generate an ensemble of distinct point patterns and then construct the corresponding Delaunay, Gabriel, Voronoi, and Delaunay-centroidal networks. For our comparison with RSA point patterns, we use the packing fraction $\phi = 0.53$ instead of the
 disorder strength $a$ as the point-pattern parameter.

We specify the ensemble sizes and plotted statistics in the relevant figure captions and discussions.

\subsection{Network analysis}\label{subsec:NetAnalysis}

\subsubsection{Ollivier--Ricci curvature (ORC)}\label{subsubsec:ORcurv}

Calculating ORC measures the local network geometry of
each edge of a network by comparing the cost of transporting probability mass between the neighborhoods of the nodes that are attached
to it~\cite{ollivier2009,Ollivier2010,Azarhooshang2020,ollivier2007,lin2011}. Let $d_G(i,j)$ denote the shortest-path distance between nodes $i$ and $j$ in a network $G$. We equip each node $i$ with the probability measure
\begin{equation} \label{orc-calc}
  m_i(z) = \begin{cases}
    			\displaystyle \frac{w_{iz}}{\sum_{z' \in \mathcal N(i)} w_{iz'}} \, , & z \in \mathcal N(i) \\[6pt]
   			 0 \, , & \text{otherwise} \, ,
		  \end{cases}
\end{equation}
where $\mathcal N(i)$ is the set of all nodes that are adjacent (i.e., connected directly) to node $i$ (not including $i$ itself) and $w_{iz}$ is the weight of the edge $(i,z)$.
In our ORC calculations,
the edge weights are $w_{ij} = 1/\ell_{ij}$ and we calculate
the shortest-path distance $d_G(i,j)$ using the
Euclidean edge lengths $\ell_{ij}$.

For nodes $i$ and $j$, the $1$-Wasserstein distance~\cite{Villani2009} between the probability measures $m_i$ and $m_j$ is
\begin{equation}\label{eq:W1}
  	W_1\bigl(m_i,m_j\bigr) = \inf_{\Pi \in \Gamma(m_i,m_j)}\sum_{u,v}\Pi(u,v)\,{d_G(u,v)} \, ,
\end{equation}
where $d_G(u,v)$ is the length of a shortest path between nodes $u, v \in G$ and $\Gamma(m_i,m_j)$ is the set of transport plans from $m_i$ to $m_j$.
A ``transport plan'' $\Pi \in \Gamma(m_i,m_j)$ is a joint probability measure over pairs of nodes, with $\Pi(u,v)$ indicating
the amount of probability mass that is transported from node $u$ to node $v$~\cite{Villani2009}. Its source-node and target-node marginal probability measures are $m_i$ and $m_j$, respectively.
The ORC of the edge $(i,j)$ is
\begin{equation}\label{eq:kappa}
 	 \kappa(i,j) = 1 - \frac{W_1\bigl(m_i,m_j\bigr)}{{d_G(i,j)}}\, .
\end{equation}
Edges with $\kappa \approx 1$ indicate redundancies in
network connectivity, whereas edges with very negative values of $\kappa$ are
bottlenecks to transport. In Sec.~\ref{sec:OR}, we compute the ORCs $\kappa(e) = \kappa(i,j)$ of the edges $e = (i,j)$ of our networks and analyze the resulting ORC distributions to compare how our network-generation methods and underlying point patterns affect local geometry.

\subsubsection{Total effective resistance (TER)}\label{subsubsec:Reff}

One can quantify a network's transport efficiency using the lens of electrical resistance~\cite{ChandraEtAl1989} by treating
a network as an electrical circuit and viewing its edges as resistors. In this analogy, the total effective resistance (TER) $\mathcal R_{\mathrm{tot}}$ of a network measures how easily current (or some other
quantity) can flow through the network. A smaller TER indicates better overall connectivity and more efficient transport, as it implies that there are more redundant paths between nodes and fewer bottlenecks along those paths \cite{Ellens2011}.
 Therefore, computing TER is useful for comparing different network structures using a single scalar value.
 Additionally,
 one can compare the robustness of networks by evaluating how their TERs
change in response to component failures \cite{Wang2014}. In the context of network design, one can treat TER as an objective function to minimize to obtain an efficient network
structure for transport~\cite{Ghosh2008}.

We calculate TER from the combinatorial graph Laplacian matrix $\mathbf{L}$~\cite{MasudaPorterLambiotte2017}, which encodes a network's connectivity structure. The entries of
$\mathbf{L}$ are
\begin{equation} \label{eq:comb_L}
  	L_{ij} = \begin{cases}
  			  \displaystyle \sum_{k} w_{ik} \, , & i = j\\
  			  -w_{ij} \, , & i \neq j \text{ and }(i,j) \in E\\
  			  0 \, , & \text{otherwise} \, .
  		\end{cases}
\end{equation}
Let $\mathbf L^{+}$ denote the Moore--Penrose pseudoinverse of $\mathbf L$ \cite{KelathayaBapatKarantha2023}, and let $\mathbf e_i$ and $\mathbf e_j$ denote unit vectors (with $1$ in the $i$th and $j$th entries, respectively, and $0$ in all other entries) that select
corresponding nodes.
The pairwise (i.e., dyadic) effective resistance between nodes $i$ and $j$ is
\begin{equation}
  	R_{ij} = \bigl(\mathbf e_i - \mathbf e_j\bigr)^{\!\mathsf T}\mathbf L^{+}\bigl(\mathbf e_i - \mathbf e_j\bigr)\, ,
\end{equation}
and the TER is
\begin{equation} \label{ter-eq}
  \mathcal R_{\mathrm{tot}} = \frac{1}{2}\sum_{i,j} R_{ij} = n\,\mathrm{Tr}\bigl(\mathbf L^{+}\bigr)\, ,
\end{equation}
where $n$ is the number of nodes and $\mathrm{Tr}$ denotes the trace of a matrix.

\subsubsection{Random‑walk mixing time}\label{subsubsec:Mixing}

For a standard random walk on a weighted, undirected network
with node set $\{1,\ldots ,n\}$, let $w_{ij} = w_{ji} \ge 0$ be the weight of edge $(i,j)$ and define the strength (i.e., weighted degree) of node $i$ by $d_i = \sum_{j = 1}^n w_{ij}$.
The transition matrix of a standard random walk is $\mathbf P = \mathbf D^{-1}\mathbf A = \mathbf I - \mathbf D^{-1}\mathbf L$, where $\mathbf A = (w_{ij})$ is the weighted adjacency matrix, $\mathbf D = \mathrm{diag}(d_1,\ldots ,d_n)$ is the diagonal matrix of strengths, and $\mathbf L = \mathbf D - \mathbf A$ is the combinatorial graph Laplacian~\cite{LevinPeresWilmer2017,MasudaPorterLambiotte2017}. See Eq.~\eqref{eq:comb_L} for the entries of $\mathbf L$.
The stationary distribution of $\mathbf P$ has entries
\begin{equation}
	\pi_i = \frac{d_i}{\sum_{k = 1}^n d_k}\, .
\end{equation}
 The total‑variation (TV) distance after $t$ steps from a starting node $i$ is
\begin{equation}
 	 \bigl\|\mathbf P^{t}(i,\cdot) - \boldsymbol\pi\bigr\|_{\mathrm{TV}} = \frac{1}{2}\sum_{j}\bigl|\mathbf P^{t}(i,j) - \pi_j\bigr| \, ,
\end{equation}
and the \emph{$\epsilon$‑mixing time}~\cite{LevinPeresWilmer2017,montenegro2006} is
\begin{equation}
 	 \tau_{\mathrm{mix}}(\epsilon) = \min\Bigl\{t:\,\max_{i}\bigl\|\mathbf P^{t}(i,\cdot) - \boldsymbol\pi\bigr\|_{\mathrm{TV}} \le \epsilon\Bigr\}\, ,
\end{equation}
where $\epsilon > 0$ is the TV-distance tolerance. For each distinct network and each discrete time $t$, we evaluate the TV distance from every starting node and take the maximum of these distances. We then determine the smallest time $t$ for which this maximum is at most $\epsilon$. Therefore, our calculation uses the least favorable starting node.

A common convention is to use $\epsilon = 1/4$ \cite{LevinPeresWilmer2017}, but we adopt a stricter convergence criterion to reduce the variance in our numerical computations. In Sec.~\ref{sec:Global}, we calculate $\tau_{\mathrm{mix}}(\epsilon)$ for $\epsilon = 10^{-3}$ for our networks and thereby corroborate the transport picture that we infer from our calculations of
ORCs and TERs.

\subsubsection{Robustness}\label{subsubsec:Robustness}

We track the evolution of the size (i.e., number of nodes) of the largest connected component (LCC) of a network relative to its total number of nodes as we progressively remove edges from it. We consider a network to be more robust than another network if its LCC retains a larger fraction of its nodes for the same fraction of removed edges~\cite{Schwarze2024, Artime2024}.

We assess network robustness by examining the change in the LCC size
as we sequentially remove edges using different strategies. We consider two edge-removal strategies: (1) random removal, in which we remove edges uniformly at random; and (2) targeted removal, in which we remove edges in the order of their ORC values, starting with the most-negative (i.e., most bottleneck-like) edges. After each edge removal, we recompute the LCC size and track it
(relative to the overall network size) as a function of the fraction of edges that we have removed. All of our statistics are means of $20$ distinct networks.
For each network, we calculate ORCs before removing any edges and use the associated edge order
throughout the targeted-removal calculation.

\subsection{Network hyperuniformity}\label{subsec:NetworkHU}

Following Maher and Newhall~\cite{maher_newhall_network_hyperuniform}, we characterize the hyperuniformity of our networks by calculating the spectral densities
of two-phase representations
of them~\cite{torquato_hyperuniform_2018}.
  To construct such a representation, we thicken each edge $(i,j)$
  of Euclidean length $\ell_{ij}$ into a rectangular beam of width $2\delta$ and length $\ell_{ij} + 2\delta$ that is centered at that edge.  We use $\delta = \rho^{-1/2}/20$, where $\rho = N_{\mathrm p}/L^2$ is the number density of the
  point pattern. These beams constitute the network phase, and the empty space between them constitutes the
  void phase.

For a network realization (i.e., a distinct network) $r$, let $\mathcal{I}{^{(r)}}(\mathbf{x})$ denote the indicator function of the network phase in a periodic simulation box of area $V = L^2$ (so $\mathcal{I}{^{(r)}}(\mathbf{x}) = 1$ if $\mathbf{x}$ is in the network phase and $\mathcal{I}{^{(r)}}(\mathbf{x}) = 0$ if $\mathbf{x}$ is in the
void phase),
and let $\phi{^{(r)}}$ denote the fraction of the area that is in the network phase.
The autocovariance function is ${\chi_{V}{^{(r)}}}(\mathbf{r}) = \langle \mathcal{I}{^{(r)}}(\mathbf{x}) \,\mathcal{I}{^{(r)}}(\mathbf{x} + \mathbf{r})\rangle - (\phi{^{(r)}})^2$, and its Fourier transform is the spectral density
\begin{equation}\label{eq:network_spectral_density}
    {{\tilde{\chi}_{V}{^{(r)}}}(\mathbf{k}) = \frac{1}{V}\left|\int_V \left[\mathcal{I}{^{(r)}}(\mathbf{x}) - \phi{^{(r)}}\right] e^{-i\mathbf{k} \cdot \mathbf{x}}\,d\mathbf{x}\right|^2 \, ,}
\end{equation}
where $\mathbf{k} = (k_x, k_y)$ denotes the wave vector.
We rasterize each two-phase representation on a $4096 \times 4096$ grid and include the square-pixel form factor in the spectral density.
The square-pixel form factor adjusts the spectral-density calculation so that we treat each occupied grid cell
as a finite square rather than as a point. For pixels of side length $\Delta = L/4096$,
this multiplicative factor is
$\operatorname{sinc}^2(k_x\Delta/2)\operatorname{sinc}^2(k_y\Delta/2)$, where $\operatorname{sinc}(z) = \sin(z)/z$ and $\operatorname{sinc}(0) = 1$~\cite{shi_keeney_chen_jiao_torquato_2023}.
To obtain one-dimensional spectra of
the two-phase network representations, we average
\eqref{eq:network_spectral_density}
over annular wave-number bins
and thereby obtain an
orientationally-averaged spectral density $\tilde{\chi}_{V,\mathrm{rad}}^{(r)}(k)$ as a function of the scalar wave number $k = |\mathbf{k}|$. Following Maher and Newhall~\cite{maher_newhall_network_hyperuniform}, we use the bin width $k_{\min}/2$, where $k_{\min} = 2\pi/L$ is the smallest accessible nonzero wave number in our periodic box.
For each combination of network-generation method, point-pattern type, and parameter value,
we calculate the mean orientationally-averaged spectral
density
\begin{equation}\label{eq:mean_radial_spectral_density}
    {\overline{\tilde{\chi}}_{V,\mathrm{rad}}(k) = \frac{1}{M}\sum_{r = 1}^{M}\tilde{\chi}_{V,\mathrm{rad}}^{(r)}(k)}
\end{equation}
of $M = 200$ distinct networks.
Using this characterization, we say that a network is ``hyperuniform" if
$\overline{\tilde{\chi}}_{V,\mathrm{rad}}(k)$ $\rightarrow 0$ as $k \rightarrow 0$~\cite{torquato_hyperuniform_2018}.

We summarize the small-$k$ behavior of a network by calculating the hyperuniformity index \cite{chen_lomba_torquato_2018}
\begin{equation}
    {H = \frac{{\overline{\tilde{\chi}}_{V,\mathrm{rad}}(0)}}{{\overline{\tilde{\chi}}_{V,\mathrm{rad}}(k_{\mathrm{peak}})}} \, .}
\end{equation}
To obtain the numerator $\overline{\tilde{\chi}}_{V,\mathrm{rad}}(0)$, we
fit a cubic polynomial to the mean orientationally-averaged spectral density at the 32 populated bins in a
small-$k$ interval.
The centers of these bins occur at the points
$k/(2\pi\sqrt{\rho}) \in [k_{\mathrm{bin,min}}/(2\pi\sqrt{\rho}), 0.30]$, where
$k_{\mathrm{bin,min}}/(2\pi\sqrt{\rho}) = (1 + \sqrt{2})/(2L\sqrt{\rho}) \approx 0.0201$. We use the polynomial-extrapolation approach of Chen
\emph{et al.}~\cite{chen_lomba_torquato_2018} and
take the fit intercept at $k = 0$.
 We define $k_{\mathrm{peak}} = \operatorname*{arg\,max}_{k > 0}\overline{\tilde{\chi}}_{V,\mathrm{rad}}(k)$
 and use the corresponding peak value
 $\overline{\tilde{\chi}}_{V,\mathrm{rad}}(k_{\mathrm{peak}})$
 as the denominator.

 For each combination of network-generation method, point-pattern type, and parameter value,
 we calculate one value of $H$ from the mean orientationally-averaged spectral density from 200 distinct networks, which we construct from 200 distinct point patterns.
 This aggregation procedure differs from the one that we use for all other network measures. (For the other measures, we first calculate one scalar value for each distinct network
 and then compute the
 mean of those
 values).
 We require that
 $H \leq 10^{-2}$ for effective hyperuniformity~\cite{maher_newhall_network_hyperuniform,chen_lomba_torquato_2018}, and we refer to the value $H = 10^{-2}$ as the ``effective-hyperuniformity threshold".

\section{Calculations of global measures}\label{sec:Global}

We now calculate mean normalized TER and mean random-walk mixing time as functions of the disorder strength $a$ of the HuPPI networks, and we quantify effective network hyperuniformity using
the hyperuniformity index $H$.

\subsection{Total effective resistance (TER) $\mathcal{R}_{\mathrm{tot}}$}\label{TER-calc}

\begin{figure}[t]
  \centering
  \includegraphics[width=\linewidth]{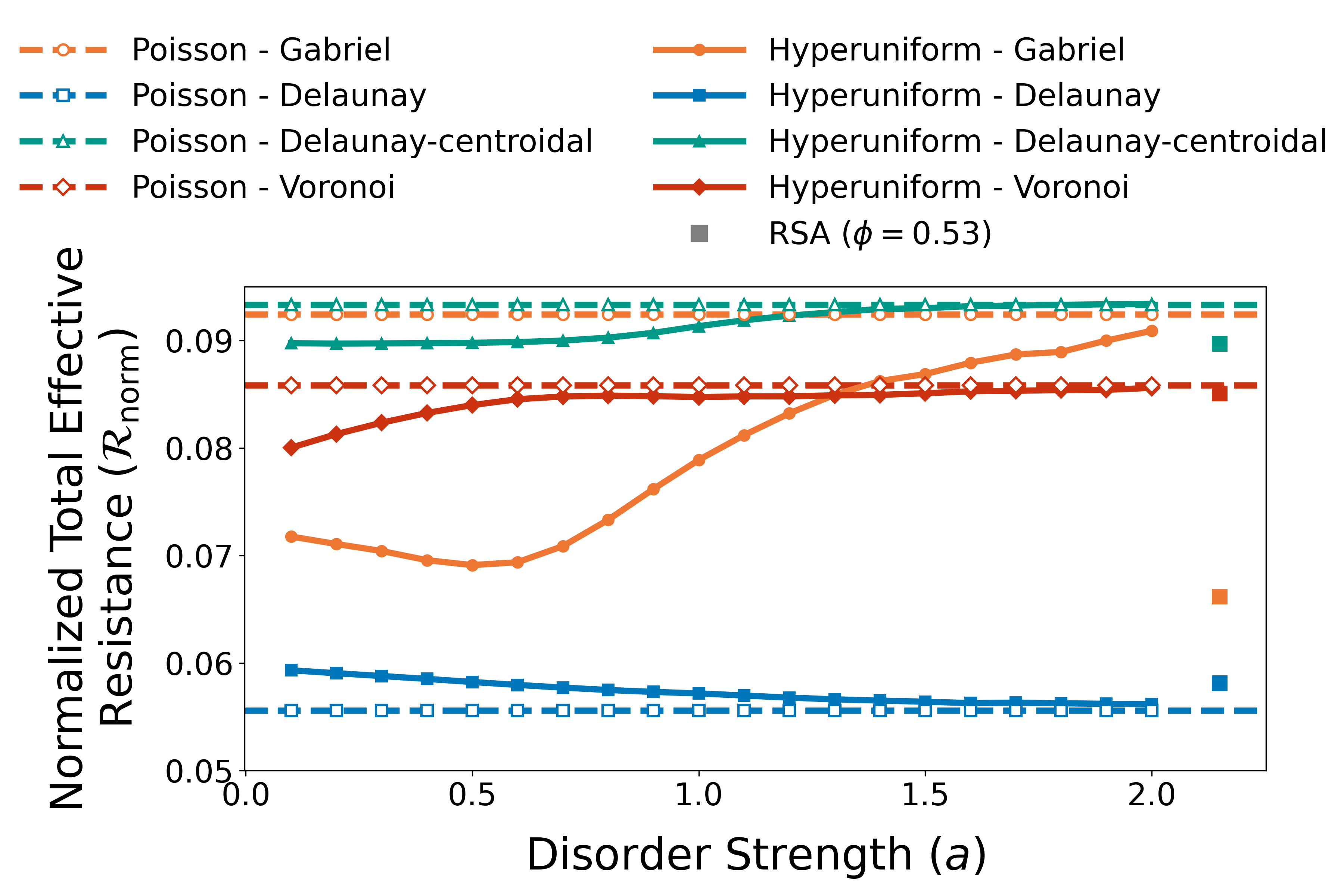}
  \caption{Mean normalized total effective resistance (TER) $\mathcal{R}_{\mathrm{norm}}$ for HuPPI networks (solid curves) as a function of disorder strength $a$, PoPPI networks (dashed lines), and networks that we generate from RSA packings with packing fraction $\phi = 0.53$ (solid square markers on the right side of the plot) in Gabriel, Delaunay, Delaunay-centroidal, and Voronoi networks. We generate each network from a point pattern with $15 \times 15 = 225$ points. For each network-generation method, we construct 100 HuPPI networks for each disorder strength $a$, 100 PoPPI networks, and 100 RSA networks. By definition, the PoPPI and RSA networks do not depend on $a$. Smaller normalized TERs indicate the presence of more redundant paths and hence indicate better connectivity.}
  \label{fig:TER_alpha}
\end{figure}

\begin{figure}[t]
  \centering
  \includegraphics[width=\linewidth]{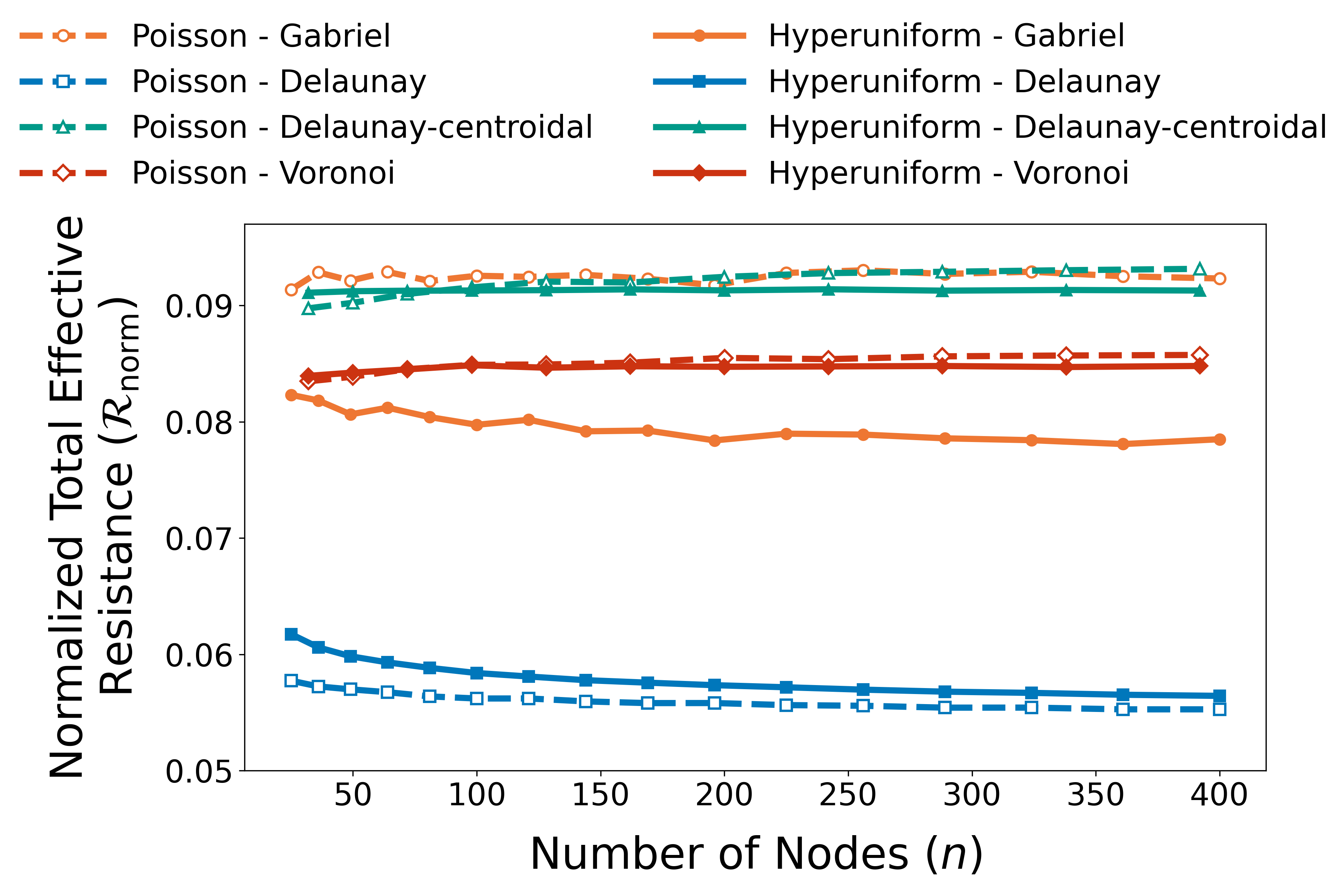}
  \caption{Mean normalized TER $\mathcal{R}_{\mathrm{norm}}$ versus the network size (i.e., number of nodes) $n$ for HuPPI networks (solid curves) and PoPPI networks (dashed curves) in Gabriel, Delaunay, Delaunay-centroidal, and Voronoi networks. The HuPPI networks have disorder strength $a = 1$. We generate each network from a point pattern with $N_{\mathrm p} = m^2$ points, with $m \in \{5, 6, \ldots, 20\}$ for Gabriel and Delaunay networks and $m \in \{4, 5, \ldots, 14\}$ for Delaunay-centroidal and Voronoi networks. [See Sec.~\ref{subsec:NetworkGen} for our discussion of the relationship between the point count (i.e., system size) $N_{\mathrm p}$ and the network size $n$.] For each network-generation method and each network size, we construct 100 HuPPI networks and 100 PoPPI networks. As we increase $n$, in Delaunay-centroidal and Voronoi networks, it becomes easier to distinguish between the mean normalized TERs in HuPPI and PoPPI networks.}
  \label{fig:TER_n}
\end{figure}

It is convenient to normalize the TER $\mathcal{R}_{\mathrm{tot}}$ by defining
\begin{equation}
	\mathcal{R}_{\mathrm{norm}} = \frac{\mathcal{R}_{\mathrm{tot}}}{n^2 \ln n}
\end{equation}
because the factor $n^{2}\ln n$ is the leading‐order growth of the TER for a square-lattice
network with periodic boundary conditions~\cite{EcevitBoysal2025,Ye01062011}. Although our
networks
are not square lattices, they are embedded in a 2D space with periodic boundary conditions.
In our empirical calculations, we observe that scaling $\mathcal{R}_{\mathrm{tot}}$ by $n^{2}\ln n$ results in a value that is approximately constant across network sizes and types.
(Simpler scaling choices do not yield a value that is approximately independent of network size.)
 In particular, we observe that $\mathcal{R}_{\mathrm{norm}}$ behaves as an intensive transport measure for the examined networks.

In Figs.~\ref{fig:TER_alpha} and \ref{fig:TER_n}, we illustrate how mean normalized TER $\mathcal{R}_{\mathrm{norm}}$ varies with the disorder strength $a$ and the network size $n$.
In Gabriel, Delaunay-centroidal, and Voronoi networks, the HuPPI networks have smaller mean normalized TERs than the PoPPI networks at most of the examined values of $a$ and $n$. In Delaunay networks, the HuPPI values are larger for the entirety of
the examined ranges.
The smaller mean normalized TERs of HuPPI networks reflect their more uniform spatial coverage.
Poisson point patterns can generate large
spatial regions with no nodes or edges.
When there are such regions, current that flows between many source--target node pairs has to detour along longer paths, and the resulting larger dyadic effective resistances accumulate to inflate the
TER $\mathcal{R}_{\mathrm{tot}}$.

The Voronoi and Delaunay-centroidal networks tend to have larger values of $\mathcal{R}_{\mathrm{norm}}$ than the
Gabriel and Delaunay networks because the
tree-like local geometries in the former two types of networks limit their numbers of redundant paths.
In Delaunay, Delaunay-centroidal, and Voronoi networks, the total number of edges is fixed by the system size $N_{\mathrm p}$ for the examined point patterns. By contrast, Gabriel networks with the same
system size
can have different numbers of edges. The Gabriel HuPPI networks at disorder strength $a = 1$ have a mean of $515.83$ edges, whereas the Gabriel PoPPI networks have a mean of $450.40$ edges. Across the 20 examined values of disorder strength,
 the Pearson correlation coefficient between the mean number of edges and the mean normalized TER of the Gabriel HuPPI networks is $r \approx -0.997$. This negative correlation indicates that Gabriel networks with more edges almost always have smaller mean normalized TERs.
 Additionally, the range of the mean normalized TER across the examined values of $a$ is
 larger for the Gabriel HuPPI networks than for the other three types of HuPPI networks.

As a comparison, we also compute TERs for networks that we generate from 100 distinct RSA packings with packing fraction $\phi = 0.53$
(see Fig.~\ref{fig:TER_alpha}). RSA packings are not hyperuniform~\cite{TorquatoUcheStillinger2006}, but the constraint that particles do not overlap with each other
introduces local order~\cite{TorquatoUcheStillinger2006,torquato2026}.
As with our HuPPI networks, the RSA networks have smaller mean normalized TERs than the PoPPI networks in Gabriel, Delaunay-centroidal, and Voronoi networks and larger mean normalized TERs than the PoPPI networks in Delaunay networks.

\subsection{Random-walk mixing time}

\begin{figure}[t]
  \centering
  \includegraphics[width=\linewidth]{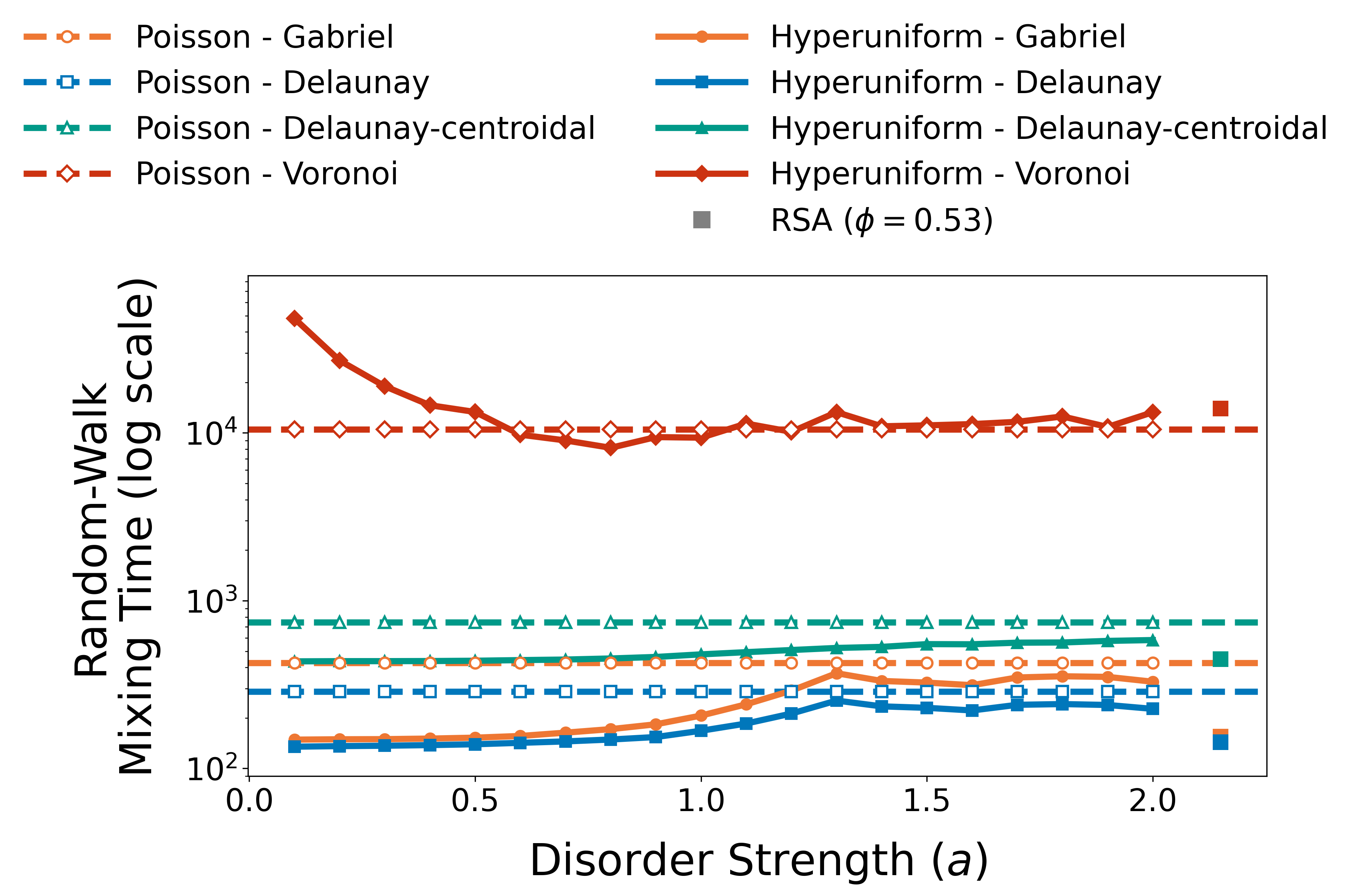}
  \caption{Mean random-walk mixing time for HuPPI networks (solid curves) as a function of disorder strength $a$, PoPPI networks (dashed lines), and RSA networks with packing fraction $\phi = 0.53$ (solid square markers on the right side of the plot) in Gabriel, Delaunay, Delaunay-centroidal, and Voronoi networks. We generate each network from a point pattern with $15 \times 15 = 225$ points. For Gabriel, Delaunay, and Delaunay-centroidal networks, we construct 100 HuPPI networks for each disorder strength $a$ and 100 PoPPI networks. For Voronoi networks, we construct 1000 HuPPI networks for each disorder strength $a$ and 1000 PoPPI networks. For each network-generation method, we construct 100 RSA networks. By definition, the PoPPI and RSA networks do not depend on $a$.}
  \label{fig:mixing}
\end{figure}

We also measure how fast a standard random walk on a network approaches its stationary distribution. We quantify this convergence speed by calculating the $\epsilon$-mixing time $\tau_{\mathrm{mix}}(\epsilon)$, which we recall (see Sec.~\ref{subsubsec:Mixing}) is the minimum discrete time $t$ such that the TV distance between a walk's distribution after $t$ steps and the stationary distribution is at most $\epsilon$. Shorter random-walk mixing times indicate more efficient linear diffusion in a network. In Fig.~\ref{fig:mixing}, we show the mean random-walk mixing times for each network-generation method for HuPPI networks, PoPPI networks, and RSA networks.

In Gabriel, Delaunay, and Delaunay-centroidal networks, the mean random-walk mixing times of the HuPPI networks are
shorter than those of their PoPPI counterparts.
In these
networks, the mean random-walk mixing times in the RSA networks are closer to the mean HuPPI mixing times than to the mean PoPPI mixing times.
Moreover, the choice of network-generation method often has a larger effect than the difference between HuPPI and PoPPI networks.
For instance, a dense Delaunay network can have shorter random-walk mixing times than a sparse Gabriel network for both hyperuniform and Poisson point patterns.

For Voronoi networks, the HuPPI networks have substantially longer mean random-walk mixing times than the PoPPI networks for small disorder strengths $a$. The mean HuPPI mixing time decreases rapidly as we increase $a$, approaches the values for PoPPI and RSA networks, and remains close to them for $a \gtrapprox 0.6$.
For small disorder strengths $a$, the Voronoi HuPPI networks have extremely short edges along cell
boundaries. Because an edge of length $\ell_{ij}$ has weight $w_{ij} = 1/\ell_{ij}$, short edges have much larger weights than other edges that are attached to their two incident nodes.
Therefore, a walker at one of these nodes is likely to traverse a short edge repeatedly instead of following an edge
that leads away from those nodes, delaying convergence to a stationary distribution. As we increase $a$, short edges become less common,
and accordingly the mean random-walk mixing time decreases.

\subsection{Network hyperuniformity}\label{subsec:NetworkHUResults}

{We now apply the two-phase spectral-density characterization from Sec.~\ref{subsec:NetworkHU} to examine how effective network hyperuniformity depends on the network-generation method and disorder strength. In Fig.~\ref{fig:network_hu}, we plot the hyperuniformity index $H$ versus the disorder strength $a$ for our four types of HuPPI networks and compare these values to those for their corresponding PoPPI networks
and to those for the RSA networks with packing fraction $\phi = 0.53$.

\begin{figure}[t]
  \centering
  \includegraphics[width=\linewidth]{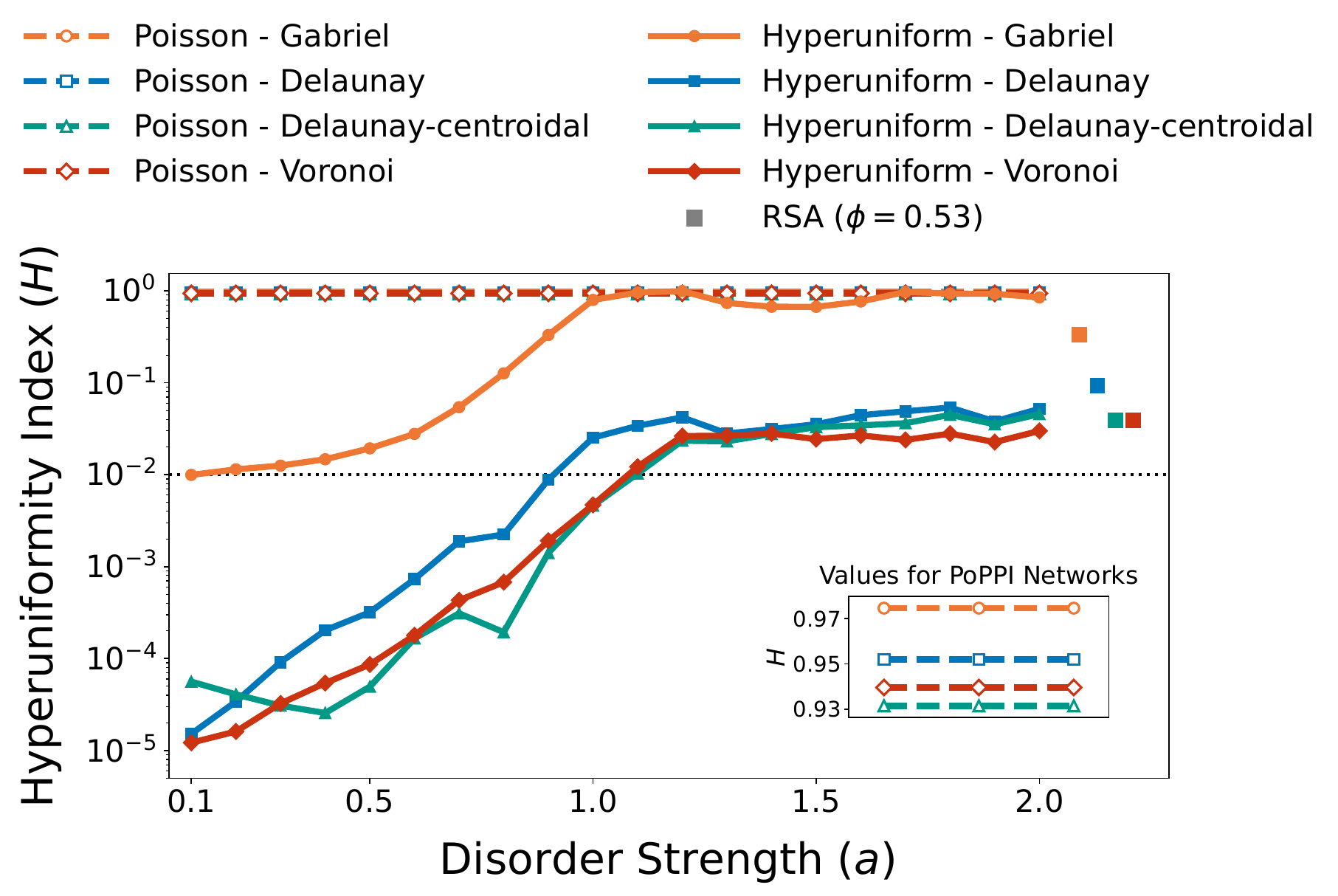}
  \caption{Hyperuniformity index $H$ for HuPPI networks (solid curves) as a function of disorder strength $a$, PoPPI networks (dashed lines), and RSA networks with packing fraction $\phi = 0.53$ (solid square markers on the right side of the plot, with horizontal offsets for visibility) in Gabriel, Delaunay, Delaunay-centroidal, and Voronoi networks. We generate each network from a point pattern with $60 \times 60 = 3600$ points. We calculate each value of $H$ for the HuPPI networks using the mean orientationally-averaged spectral density $\overline{\tilde{\chi}}_{V,\mathrm{rad}}(k)$ from 200 distinct networks, which we construct from 200 distinct point patterns at each disorder strength $a$. Analogously, we calculate each value of $H$ for the PoPPI networks and RSA networks using the mean orientationally-averaged spectral density from 200 distinct networks (which we also construct from 200 distinct point patterns). By definition, the PoPPI and RSA networks do not depend on $a$. The PoPPI networks have distinct but very similar values of $H$, which range from $H \approx 0.93$ to $H \approx 0.97$, so their associated dashed lines almost coincide. In the inset, we show magnifications of the values of $H$ for the PoPPI networks on a linear scale. The black dotted line in the main plot marks the effective-hyperuniformity threshold $H = 10^{-2}$.}
  \label{fig:network_hu}
\end{figure}

At disorder strength $a = 0.1$, the value of the hyperuniformity index $H$ for the Gabriel HuPPI networks
is $H \approx 9.95 \times 10^{-3}$, which is near the effective-hyperuniformity threshold $H = 10^{-2}$.
For the Gabriel HuPPI networks, the values of $H$ exceed the effective-hyperuniformity threshold when $a \geq 0.2$.
The Delaunay HuPPI networks remain effectively hyperuniform
through $a = 0.9$, and their values of $H$ cross the
threshold $H = 10^{-2}$ between $a = 0.9$ and $a = 1$. The Delaunay-centroidal and Voronoi HuPPI networks are effectively hyperuniform when $a \leq 1$, and their values of $H$ cross the
threshold $H = 10^{-2}$ between $a = 1$ and $a = 1.1$.
For the RSA networks, $H \approx 0.331$ for Gabriel networks, $H \approx 9.35 \times 10^{-2}$ for Delaunay networks, $H \approx 3.89 \times 10^{-2}$ for Delaunay-centroidal networks, and $H \approx 3.88 \times 10^{-2}$ for Voronoi networks. All four values of $H$ exceed the effective-hyperuniformity threshold $H = 10^{-2}$, but they are substantially smaller than the values of $H$ for the corresponding PoPPI networks.

Our results are consistent with the dependence of effective hyperuniformity on network-generation method and disorder strength that
was reported in~\cite{maher_newhall_network_hyperuniform}. In particular, although the underlying hyperuniform point patterns of our HuPPI networks remain class-I hyperuniform as we increase $a$, their two-phase network representations lose effective hyperuniformity at
different disorder strengths for our four network-generation methods.
At each examined disorder strength
between $a = 1$ and $a = 2$, the value of the hyperuniformity index $H$ for the Gabriel HuPPI networks is comparable to that for their PoPPI counterparts.
At each examined disorder strength
between $a = 1.1$ and $a = 2$, the values of $H$ for the Delaunay, Delaunay-centroidal, and Voronoi HuPPI networks exceed
the effective-hyperuniformity threshold $H = 10^{-2}$ (and hence are not effectively hyperuniform) but are much smaller than
the values of $H$ for their corresponding PoPPI networks.

\begin{figure*}[t]
  \centering
  \includegraphics[width=\linewidth]{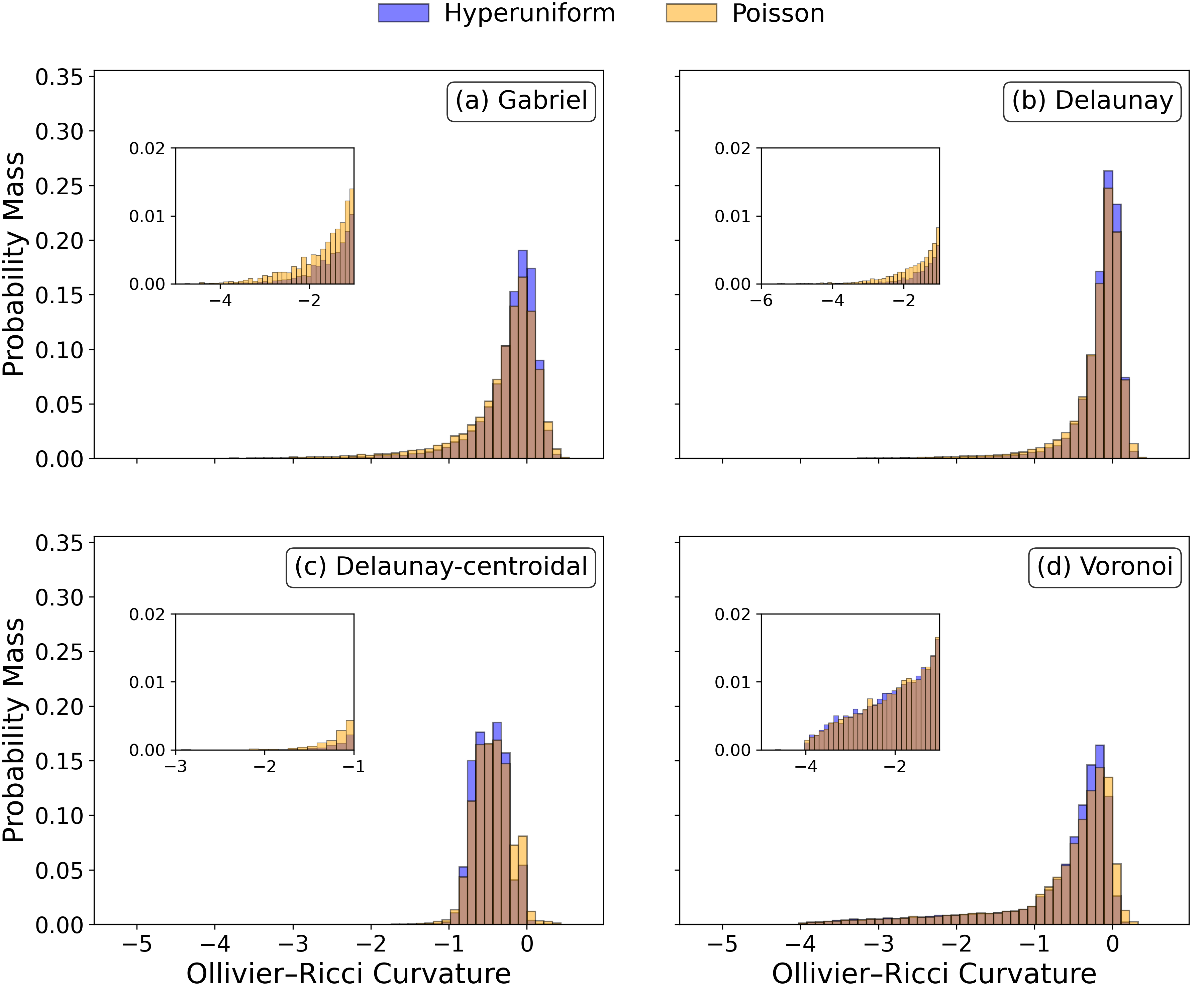}
  \caption{Histograms of the ORCs of the edges of the HuPPI networks (blue) with disorder strength $a = 1$ and PoPPI networks (orange) in Gabriel, Delaunay, Delaunay-centroidal, and Voronoi networks. In the insets, we show magnifications of the left tails of the histograms. We generate each network from a point pattern with $15 \times 15 = 225$ points. The histograms aggregate data from 50 distinct point patterns of each type.}
  \label{fig:curv_hist}
\end{figure*}

\section{Ollivier--Ricci curvature (ORC): A local measure}\label{sec:OR}

\subsection{ORC distributions}

We now examine the distribution of ORC values across edges to compare how point-pattern type and network-generation method shape local geometric order.

In Fig.~\ref{fig:curv_hist}, we show the ORC distributions for both HuPPI and PoPPI networks.
Our observations underscore the fact that the point-pattern type (hyperuniform versus Poisson point patterns) and the network type (Gabriel and Delaunay networks versus Delaunay-centroidal and Voronoi networks) jointly influence the shape of the ORC distribution.
In Gabriel and Delaunay networks, the PoPPI networks have
wider ORC distributions than their HuPPI counterparts. The difference is smaller but present in Delaunay-centroidal networks, and the ORC distributions in the HuPPI and PoPPI Voronoi networks have almost the same width.
The Delaunay networks also tend to have larger ORCs than the other types of networks.
Additionally, adjacent nodes in Voronoi and Delaunay-centroidal networks generally share few common neighbors. This limited overlap between their neighborhoods yields more-negative ORCs than in Delaunay and Gabriel networks.
We also observe that the Delaunay and Gabriel networks have similar ORC distributions. Recall that a Gabriel network is a subgraph of a Delaunay network,
so every edge of a Gabriel network is also an edge of a Delaunay network, but any Delaunay-network edge whose diametral disk has additional points is not in its associated Gabriel subgraph.
The Delaunay and Gabriel network-generation mechanisms both connect nodes
based on an empty-disk requirement (via circumscribed circles in Delaunay networks and via diametral disks in Gabriel networks), so they have similar local connectivity patterns.
Therefore, many of the edges of these networks have neighborhoods with similar local geometries.

\subsection{Mean ORC versus disorder strength $a$}

\begin{figure}[t]
  \centering
  \includegraphics[width=\linewidth]{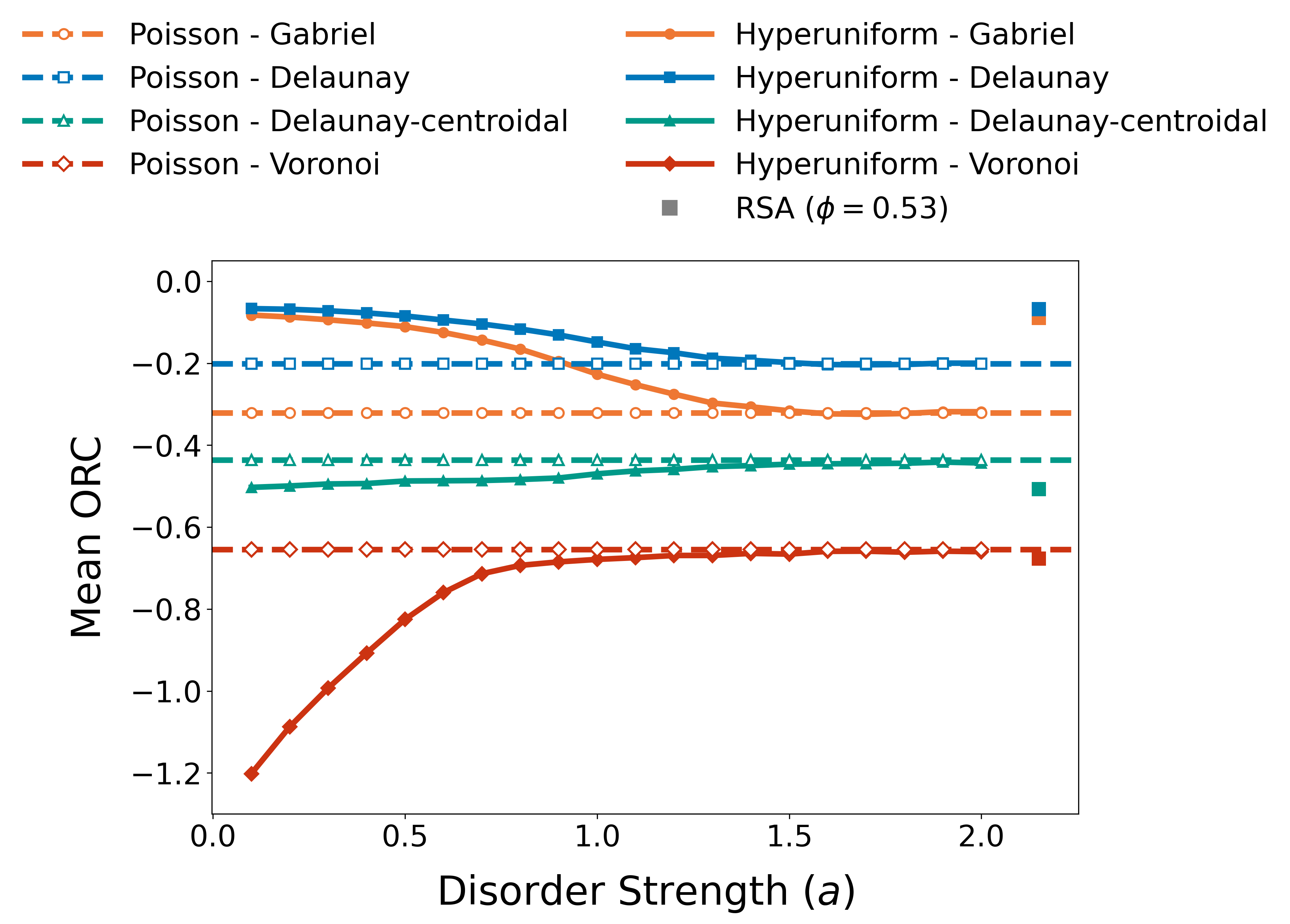}
  \caption{Mean ORC for HuPPI networks (solid curves) as a function of disorder strength $a$, PoPPI networks (dashed lines), and RSA networks with packing fraction $\phi = 0.53$ (solid square markers on the right side of the plot) in Gabriel, Delaunay, Delaunay-centroidal, and Voronoi networks. We generate each network from a point pattern with $15 \times 15 = 225$ points. For each network-generation method, we construct 50 HuPPI networks for each disorder strength $a$, 50 PoPPI networks, and 100 RSA networks. By definition, the PoPPI and RSA networks do not depend on $a$.}
  \label{fig:curv_alpha}
\end{figure}

We now examine how the disorder strength $a$ affects the mean ORC. We show our results in Fig.~\ref{fig:curv_alpha}.
As we increase $a$, the mean ORCs of the HuPPI networks approach those of the corresponding PoPPI networks, even though the underlying point patterns remain hyperuniform at the examined disorder
strengths~\cite{klatt_cloaking_2020}. In the Delaunay-centroidal and Voronoi networks, the mean ORCs of the HuPPI networks approach those of the corresponding PoPPI networks from below as we increase $a$.
In the Gabriel and Delaunay networks, the mean ORCs of the HuPPI networks approach those of the corresponding PoPPI networks from above.

We also observe in Fig.~\ref{fig:curv_alpha} that the network-generation method has a large effect on the mean ORC.
At each examined disorder strength, the Delaunay HuPPI networks have the largest mean ORCs and the Voronoi HuPPI networks have the smallest mean ORCs.
As we increase $a$, the mean ORCs of the Delaunay and Gabriel HuPPI networks decrease,
 whereas those of the Delaunay-centroidal and Voronoi HuPPI networks increase.
  In both of these pairs, the networks are related to each other.
  Delaunay and Gabriel networks have similar local connectivity patterns, and
   Delaunay-centroidal and Voronoi networks that we generate from the same point pattern differ only in their node locations (see Sec.~\ref{subsec:NetworkGen}).

As we discussed previously, Gabriel networks
can have different numbers of edges for a fixed system size $N_{\mathrm p}$ (see Sec.~\ref{subsec:NetworkGen}) and
the mean normalized TERs of Gabriel HuPPI networks are negatively correlated with their mean numbers of edges (see Sec.~\ref{TER-calc}).
Across the examined disorder strengths $a$, the Pearson correlation coefficient between the mean number of edges and the mean ORC of the Gabriel HuPPI networks is $r \approx 0.986$.
Therefore, Gabriel HuPPI networks with
larger mean numbers of edges
have less-negative mean ORCs.

In Voronoi HuPPI networks, edges that are much shorter than the other edges that are attached to their incident nodes
become less
common
as we increase the disorder strength $a$.
In our calculation of ORCs, the probability measures $m_i$ and $m_j$ (see Eq.~\eqref{orc-calc}) depend on the edge weights, whereas the graph distance $d_G$ depends on Euclidean edge lengths.
Therefore, the distance $d_G(i,j)$ is very small for short edges $(i,j)$.
In Voronoi networks, adjacent nodes typically
share few common neighbors, so the cost of transporting probability mass between their neighborhoods
is large relative to this small graph distance. Consequently, $W_1\bigl(m_i,m_j\bigr)/d_G(i,j)$ is large
and Eq.~\eqref{eq:kappa} yields a very
negative value of $\kappa(i,j)$.
As we increase the disorder strength $a$, the mean ORCs of the Voronoi HuPPI networks become less negative.
Short edges have very negative ORCs,
so the decrease in their prevalence as we increase $a$ helps explain this trend.
However,
the prevalence of short edges
does not fully explain
the difference between
the mean ORCs of
the Voronoi HuPPI and PoPPI networks
for $a \gtrapprox 1$, where short edges are almost as common in the PoPPI networks as in the HuPPI networks.

In the Gabriel, Delaunay, and Delaunay-centroidal networks, the mean ORCs of the RSA networks are close to those of the HuPPI networks
for disorder strengths between $a = 0.1$ and $a = 0.3$.
In the Voronoi networks, the mean ORCs of the RSA networks are close to those of the HuPPI networks at disorder strength $a = 1$ and are more negative than those of the PoPPI networks.
Because RSA packings are not hyperuniform, the
similarities between RSA and HuPPI networks in mean ORCs and mean normalized TERs (see Fig.~\ref{fig:TER_alpha})
demonstrate that local order impacts
the observed differences between HuPPI and PoPPI networks.}

In summary, for all four network-generation methods, the mean ORCs of the HuPPI networks become more similar to those of the corresponding PoPPI networks as we increase the disorder strength $a$. This increase in similarity is consistent with our observations in Sec.~\ref{subsec:NetworkHUResults} (and with the findings in~\cite{maher_newhall_network_hyperuniform}) that the effective hyperuniformity of a network's two-phase representation depends on the network-generation method and that the two-phase representations of HuPPI networks lose effective hyperuniformity as one increases the disorder strength $a$.

\subsection{Convergence of mean ORC with increasing network size}

\begin{figure}[t]
  \centering
  \includegraphics[width=\linewidth]{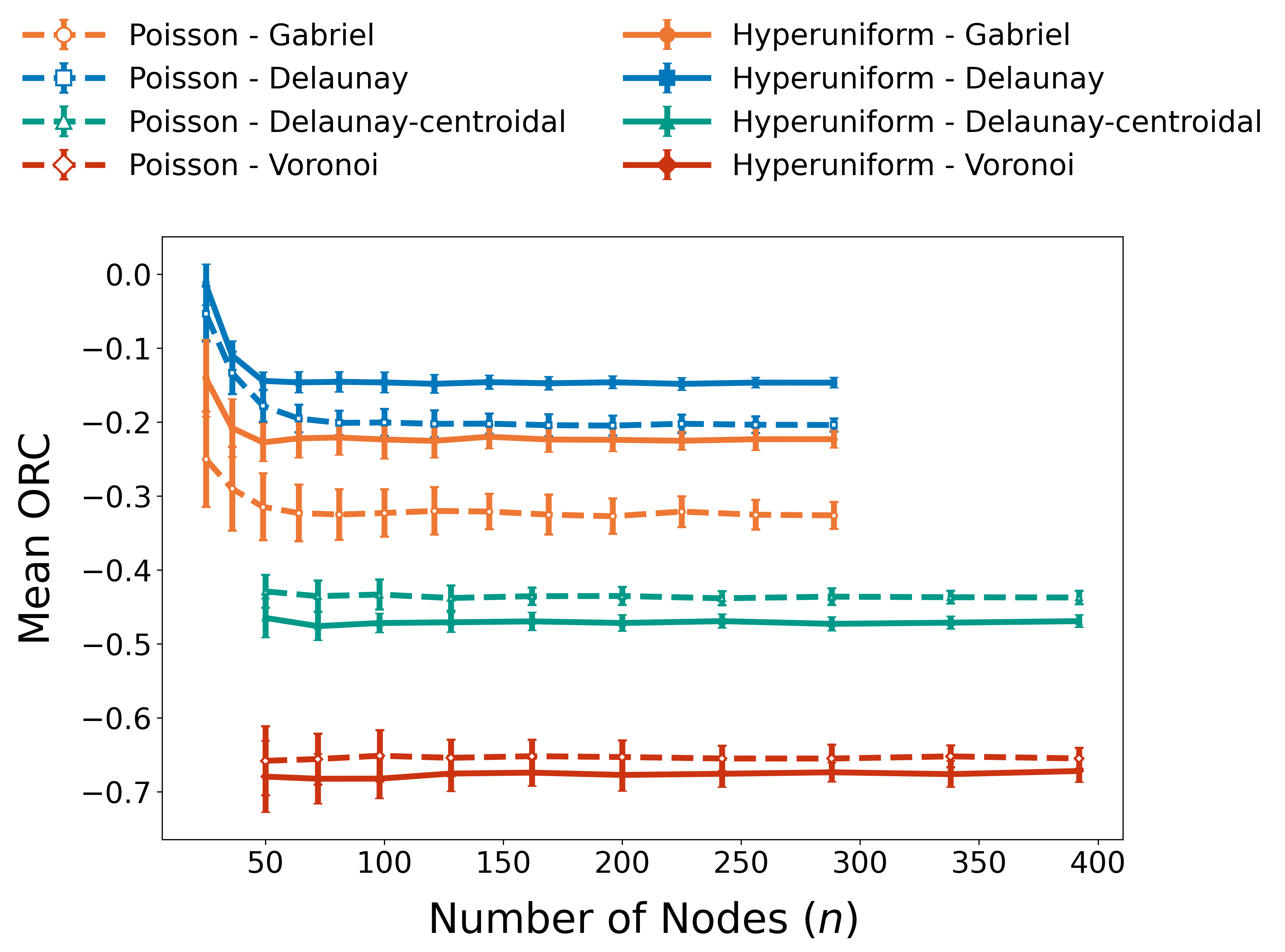}
  \caption{Mean ORC versus network size (i.e., number of nodes) $n$ for HuPPI networks (solid curves) with disorder strength $a = 1$ and PoPPI networks (dashed curves) in Gabriel, Delaunay, Delaunay-centroidal, and Voronoi networks. The curves give mean ORCs for 50 distinct networks for each combination of point-pattern type, network-generation method, and network size, with error bars indicating one standard deviation from the mean. We generate each network from a point pattern with $N_{\mathrm p} = m^2$ points, with $m \in \{5, 6, \ldots, 17\}$ for Gabriel and Delaunay networks and $m \in \{5, 6, \ldots, 14\}$ for Delaunay-centroidal and Voronoi networks. [See Sec.~\ref{subsec:NetworkGen} for our discussion of the relationship between the point count (i.e., system size) $N_{\mathrm p}$ and the network size $n$.]}
  \label{fig:curv_N}
\end{figure}

We now examine how ORC changes with the network size (i.e., number of nodes) $n$. In Fig.~\ref{fig:curv_N}, we plot the mean ORC versus $n$ in our HuPPI and PoPPI networks.
At each examined network size, the mean ORCs of the Delaunay-centroidal and Voronoi HuPPI networks are smaller than those of their PoPPI counterparts, whereas the mean ORCs of the Gabriel and Delaunay HuPPI networks are larger than those of their PoPPI counterparts.
In networks with roughly $100$ or more nodes, the mean ORCs change little as we increase $n$.
This is a promising observation for analyzing mean ORCs in physical networks that one constructs experimentally in laboratories (e.g., via additive manufacturing~\cite{Obrero2024,rock2025}), as it suggests that even modest network sizes
allow us to successfully estimate mean ORCs in the types of networks that we examine.

\subsection{Edge-level correlations between ORC and TER}\label{subsec:EdgeCorr}

\begin{figure*}[t]
  \centering
  \includegraphics[width=\linewidth]{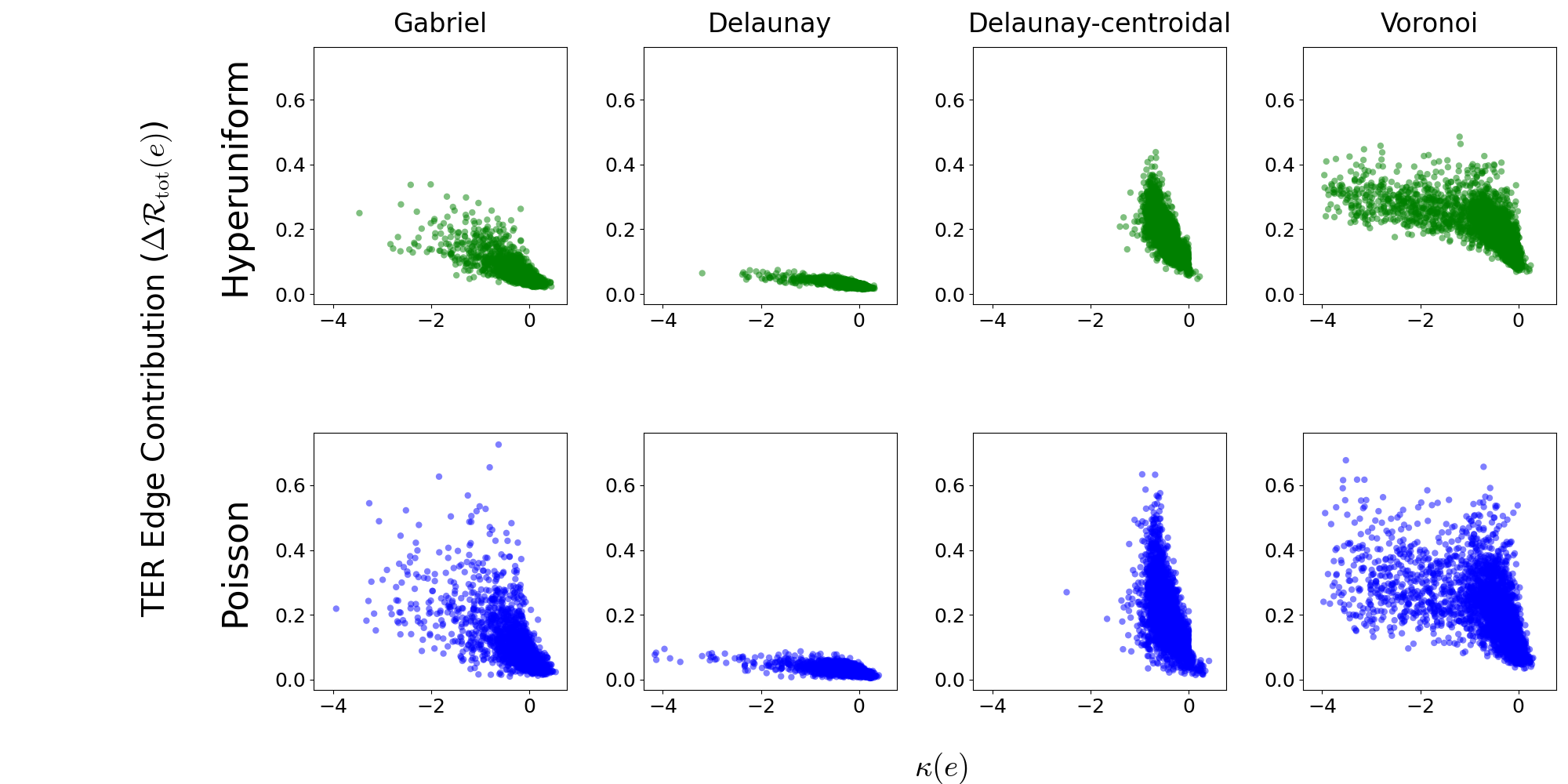}
  \caption{Scatter plots of the ORCs $\kappa(e)$ versus the TER edge contributions $\Delta \mathcal{R}_{\mathrm{tot}}(e)$ of edges $e = (i, j)$ in (top) HuPPI networks with disorder strength $a = 1$ and (bottom) PoPPI networks in Gabriel, Delaunay, Delaunay-centroidal, and Voronoi networks (which we show from left to right). In each column, we show data from a single realization of both a HuPPI network and a PoPPI network. For each network-generation method, we construct 10 HuPPI networks and 10 PoPPI networks to calculate the means and standard deviations of the Pearson correlation coefficients that we report in the main text. We generate each network from a point pattern with $33 \times 33 = 1089$ points.}
  \label{fig:edge_corr}
\end{figure*}

Comparing our
global transport results (see Sec.~\ref{sec:Global}) with the
ORC statistics reveals a counterintuitive situation. Negative ORCs typically indicate the presence of bottlenecks~\cite{Ni2015,Sandhu2016,Azarhooshang2020,Gosztolai2021}, so one may expect that networks with
more-negative mean ORCs have larger TERs.
However, at disorder strength $a = 1$, mean ORC alone does not explain the differences in mean normalized TER between HuPPI and PoPPI networks in Delaunay, Delaunay-centroidal, and Voronoi networks.
In Delaunay-centroidal and Voronoi networks, the HuPPI networks have more-negative mean ORCs (see Fig.~\ref{fig:curv_alpha}) and smaller mean normalized TERs (see Fig.~\ref{fig:TER_alpha}) than their PoPPI counterparts.
However, in Delaunay networks,
the HuPPI networks have less-negative mean ORCs and larger mean normalized TERs than their PoPPI counterparts. The Gabriel networks are the only examined networks for which the HuPPI networks have both less-negative mean ORCs and smaller mean normalized TERs than their PoPPI counterparts. At disorder strength $a = 1$, the mean number of edges is also larger in the Gabriel HuPPI networks than in their PoPPI counterparts (see Sec.~\ref{sec:Global}).
To investigate
this counterintuitive situation and gain insight into how local geometric order relates to TER, we examine
the relationship between the ORC $\kappa(e)$ of an
edge $e$ and its specific contribution to TER. We calculate the \emph{TER edge contribution}
\begin{equation}
	\Delta \mathcal{R}_{\mathrm{tot}}(e) = \mathcal{R}_{\mathrm{tot}}(G \setminus\! \{e\}) - \mathcal{R}_{\mathrm{tot}}(G) \, ,
\end{equation}
where $\mathcal{R}_{\mathrm{tot}}(G \setminus\! \{e\})$ is the TER of a network after removing edge $e$ and $\mathcal{R}_{\mathrm{tot}}(G)$ is the TER
of the original network. Our edge-removal perspective is related to Lee
\emph{et al.}'s transport-based ``path score"~\cite{lee2014}, which quantifies the importance of an edge by how often it appears on shortest ``backup paths'' after one removes other edges. By contrast, $\Delta \mathcal{R}_{\mathrm{tot}}(e)$ assigns the change in TER to a removed edge.
In Fig.~\ref{fig:edge_corr}, we show scatter plots of $\kappa(e)$ versus $\Delta \mathcal{R}_{\mathrm{tot}}(e)$ for point patterns with $N_{\mathrm p} = 1089$ points.

The relationship between the
ORC $\kappa(e)$ and the TER edge contribution $\Delta \mathcal{R}_{\mathrm{tot}}(e)$
depends on the network-generation method. For each network type, we compute the Pearson correlation coefficient $r$ between $\kappa(e)$ and $\Delta \mathcal{R}_{\mathrm{tot}}(e)$ using
10 distinct networks
and report the mean
and standard deviation.
For HuPPI networks, the correlations are $r \approx -0.775 \pm 0.012$ for Gabriel networks, $r \approx -0.773 \pm 0.012$ for Delaunay networks, $r \approx -0.745 \pm 0.008$ for Delaunay-centroidal networks, and
$r \approx -0.702 \pm 0.008$ for Voronoi networks. For PoPPI networks, the correlations are $r \approx -0.618 \pm 0.018$ for Gabriel networks, $r \approx -0.617 \pm 0.023$ for Delaunay networks, $r \approx -0.584 \pm 0.017$ for Delaunay-centroidal networks, and $r \approx -0.554 \pm 0.012$ for Voronoi networks.
These negative correlations indicate that edges with more-negative ORCs tend to have larger TER edge contributions. For each network-generation method, the magnitude of this correlation is larger in HuPPI networks than in PoPPI networks. However, the scatter of the values in Fig.~\ref{fig:edge_corr} demonstrates
that the ORC of an edge does not uniquely determine its TER edge contribution.

In Fig.~\ref{fig:edge_corr}, we also observe that the PoPPI networks have wider distributions of TER edge contributions than their HuPPI counterparts for all four network-generation methods. The horizontal spreads are consistent with the ORC distributions in Fig.~\ref{fig:curv_hist}. The difference between the horizontal spreads in the HuPPI and PoPPI networks is very noticeable in Gabriel and Delaunay networks, small (except for an outlier) in Delaunay-centroidal networks, and very small in Voronoi networks. The wider vertical distributions in the PoPPI networks than in the HuPPI networks include outlier edges with very large TER edge contributions. Importantly, the edges in PoPPI networks
with the largest TER edge contributions have negative ORCs and are critical bridges, as removing them from a network severely disrupts global transport. This observation is consistent with the interpretation of negative-ORC edges as bridge-like bottlenecks. At disorder strength $a = 1$, the maximum TER edge contributions in the HuPPI networks are always smaller than those in the corresponding PoPPI networks (see Fig.~\ref{fig:edge_corr}).
The
critical bridges elevate the values of TER $\mathcal R_{\mathrm{tot}}$ in the PoPPI networks.
Although the Delaunay-centroidal and Voronoi HuPPI networks have more-negative mean ORCs than their PoPPI counterparts at disorder strength $a = 1$, these HuPPI networks have few extreme bottleneck edges.
By contrast, extreme bottleneck edges dominate the TER in PoPPI networks.
Consequently, in Gabriel, Delaunay-centroidal, and Voronoi networks, we conclude that the smaller mean normalized TERs in HuPPI networks than in PoPPI networks arise from the scarcity of edges with large
values of $\Delta \mathcal{R}_{\mathrm{tot}}(e)$.
In Delaunay-centroidal and Voronoi networks, this scarcity explains why the HuPPI networks with disorder strength $a = 1$ have smaller mean normalized TERs than their PoPPI counterparts despite their more-negative mean ORCs.

\section{Robustness of networks to edge removals}\label{sec:Robustness}

\begin{figure*}[t]
  \centering
  \includegraphics[width=0.9\linewidth]{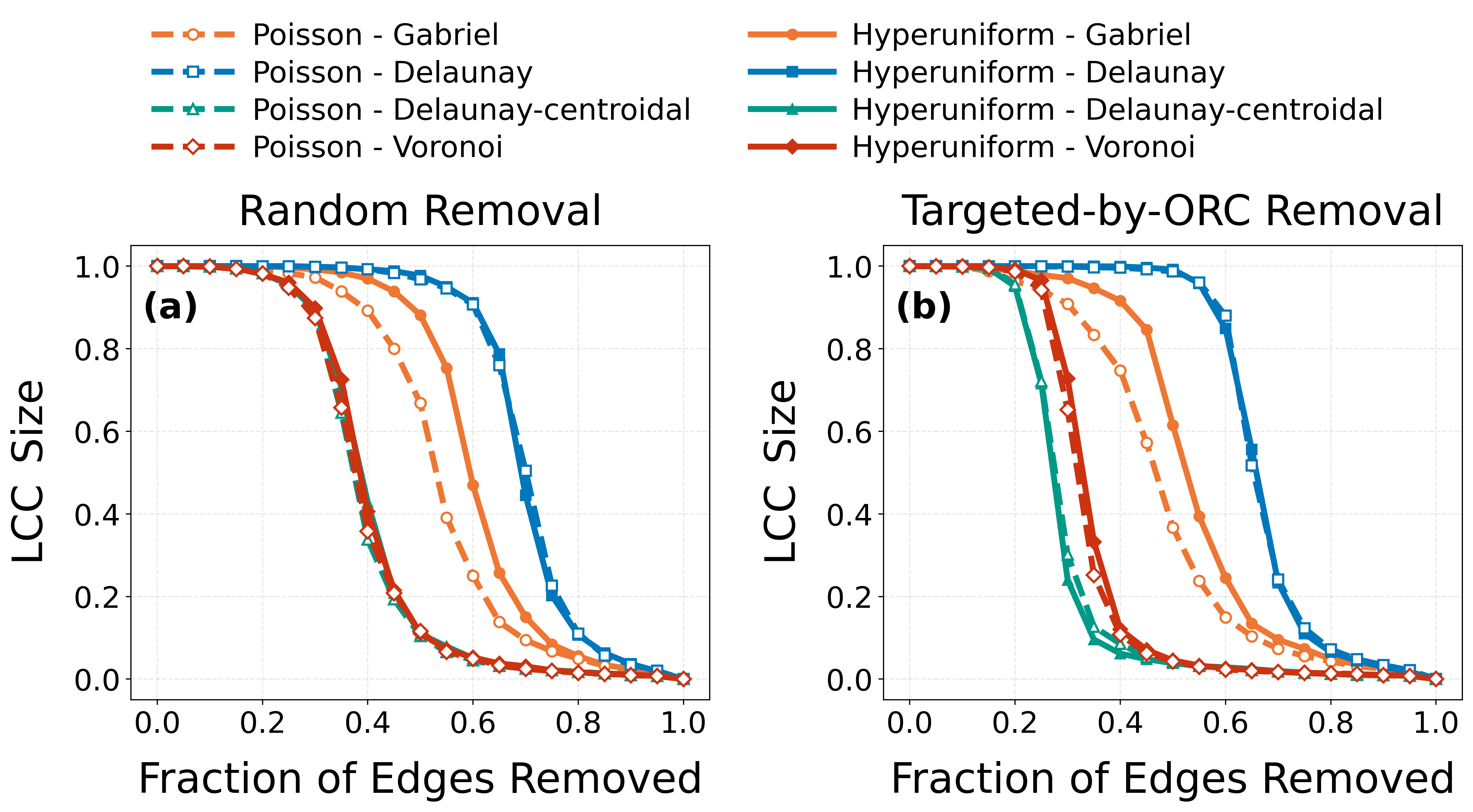}
  \caption{Mean relative sizes of the largest connected components (LCCs) of the networks versus the fraction of edges removed for two different removal strategies for HuPPI networks (solid curves) with disorder strength $a = 1$ and PoPPI networks (dashed curves) in Gabriel, Delaunay, Delaunay-centroidal, and Voronoi networks. We generate each network from a point pattern with $15 \times 15 = 225$ points. For each network-generation method, we construct 20 HuPPI networks and 20 PoPPI networks. We show results for (a) uniformly random removal of edges and (b) removal of edges in the order of their ORC values starting with the most-negative value.}
  \label{fig:robustness}
\end{figure*}

Having established that PoPPI networks possess outlier edges with very large TER edge contributions $\Delta \mathcal{R}_{\mathrm{tot}}(e)$
that correspond to critical bridges (whose removal severely disrupts global transport), we now study the robustness of HuPPI and PoPPI networks. We examine network robustness by progressively removing edges from
networks and calculating the sizes of their ensuing LCCs relative to the network sizes~\cite{Schwarze2024}.
For each network-generation method, we calculate the mean relative LCC size of 20 HuPPI networks and the mean relative LCC size of 20 PoPPI networks.
Cetinay
\emph{et al.}~\cite{dev2018} demonstrated that targeting edges (or nodes) based on their curvatures or resistance-based centralities is able to identify structural bottlenecks in power-transmission networks, and they illustrated that removing such edges significantly disrupts network connectivity. Taking inspiration from their results, we consider two edge-removal strategies: (1) random removal, in which we remove edges uniformly at random; and (2) targeted removal, in which we remove edges in the order of their ORC values, starting with the edges with the most-negative values (i.e., the most bottleneck-like edges).

For all four network-generation methods, HuPPI and PoPPI networks tend to fragment more rapidly when we remove edges in the order of their ORC values, starting with the most-negative values, than when we remove edges uniformly at random.
For both uniformly random edge removal (see Fig.~\ref{fig:robustness}(a)) and targeted edge removal (see Fig.~\ref{fig:robustness}(b)), the mean relative LCC size is larger in the Gabriel HuPPI networks than in their PoPPI counterparts. Additionally, for both edge-removal
strategies, the curves of the mean relative LCC size are very similar in the HuPPI networks and their PoPPI counterparts in Delaunay, Delaunay-centroidal, and Voronoi networks.

\section{Conclusions and discussion}\label{sec:Conc}

We studied the influence of point-pattern hyperuniformity, which is characterized by the suppression of long-range density fluctuations in comparison to the fluctuations in standard disordered systems, on the structure of spatial networks. We calculated several global and local network measures --- Ollivier--Ricci curvature (ORC) distributions, total effective resistance (TER), random-walk mixing times, and robustness to edge removal --- in spatially-embedded networks that we generated from hyperuniform point patterns, Poisson point patterns, and random-sequential-adsorption (RSA) packings. In our calculations, we compared networks that we constructed using Delaunay, Gabriel, Voronoi, and Delaunay-centroidal network-generation methods.

Our numerical computations suggest two key conclusions. First, depending on the network-generation mechanism, hyperuniform point patterns
can confer transport and robustness advantages to spatial networks.
Hyperuniform-point-pattern-induced (HuPPI) networks have smaller mean normalized TERs
than Poisson-point-pattern-induced (PoPPI) networks
in Gabriel, Delaunay-centroidal, and Voronoi networks;
have shorter mean random-walk mixing times
than PoPPI networks
in Gabriel, Delaunay, and Delaunay-centroidal networks;
and, in Gabriel networks, are more robust than PoPPI networks when we remove edges either uniformly at random or in the order of their ORC values, starting with the most-negative values.
The scarcity of critical bridges (i.e., edges with negative ORCs and very large TER edge contributions) in HuPPI networks helps explain their smaller mean normalized TERs in Gabriel, Delaunay-centroidal, and Voronoi networks.
We also observed that RSA networks sometimes have
mean normalized TERs and mean ORCs
that are similar to those of
HuPPI networks.
Because RSA packings have local order but are not hyperuniform, these similarities
demonstrate that local order impacts
the observed differences between HuPPI and PoPPI networks.
Second, we observed that the network-generation method strongly influences the examined network measures, providing an additional design lever
alongside the choice of point pattern.
In several of our comparisons of network properties, the choice of network-generation method has a larger effect on the examined transport and curvature measures than the choice between hyperuniform and Poisson point patterns.
For example, the relationship between point-pattern type and mean ORC depends on the network-generation method. In particular, we observed that (1) the Delaunay and Gabriel networks are similar to each other and that (2) the Delaunay-centroidal and Voronoi networks are similar to each other. The Delaunay and Gabriel networks have similar ORC distributions and similar local structures, and they experience similar changes in their mean ORCs as we increase
disorder strength. Additionally, the Delaunay-centroidal and Voronoi networks have similar local structures and experience similar changes in their mean ORCs as we increase disorder strength.

The mean normalized TERs, mean random-walk mixing times, and mean ORCs of the HuPPI networks
approach those of the PoPPI networks as we increase the disorder strength $a$.
These trends occur even though the underlying point patterns remain hyperuniform at the examined disorder strengths.
Despite these increasingly similar transport and curvature measurements as we increase disorder strength, the two-phase representations of the Delaunay, Delaunay-centroidal, and Voronoi HuPPI networks retain much smaller hyperuniformity indices (and hence larger effective hyperuniformities) than their PoPPI counterparts.
We thus conclude that the examined transport and curvature measures in the HuPPI networks become more similar
to those in the PoPPI networks as one increases disorder strength.

In summary, our computations
demonstrate that hyperuniform point patterns
can confer transport and robustness advantages
to spatial networks (depending on the network-generation method).
The structure of HuPPI, PoPPI, and RSA networks depends both on the correlations in their underlying point patterns and on the
network-generation method. These insights motivate both (1) further theoretical efforts to relate edge-level correlations between ORC and TER edge contribution to global transport properties
and (2) experimental studies to quantify how specific spatial point patterns combine with particular network-generation methods
to influence network performance in physical systems.

Disordered systems offer a much larger design space than ordered systems~\cite{ebra2026,wytock2026}, so it is important to examine how tuning the disorder strength in hyperuniform point patterns impacts the properties of
the resulting HuPPI networks and other disordered metamaterials. Recent experimental investigations of electrical and mechanical transport in disordered networks~\cite{Obrero2024,rock2025,Yang2025_cloaking} and computational investigations of network hyperuniformity \cite{maher_newhall_network_hyperuniform}
and the spectra of Laplacian differential operators
on HuPPI networks~\cite{maher2026spectrum} provide promising foundations for such work.

\section*{Data availability}

Our code to generate our point patterns and our networks, compute
network measurements, and generate our figures is available at \url{https://github.com/DMREF-networks/HuPPI-Network-Analysis} and is archived in the Zenodo repository~\cite{Raj2026HuPPICode}.
Our data, which consist of all point patterns and
adjacency matrices, are available in a searchable
database of
network configurations
at \url{https://dmref-networks.github.io/disordered-configurations/}.

\medskip
\medskip

\section*{Acknowledgements}

We gratefully acknowledge funding from the National Science Foundation through a collaborative grant via NSF DMREF Grant Nos. CMMI-2323343 (JVR and MAP) and CMMI-2323342 (CEM and KAN).
We thank Robijn Bruinsma, Karen Daniels, Van Savage, and an anonymous referee for helpful comments.
We used ChatGPT and Codex (with models from o3 through GPT-5.6 Sol) and
Claude (with models from Opus 4.5 through Fable 5) to assist with the development and debugging of portions of our simulation code, preparing the reference list, and checking grammar and individual sentences. We independently verified all output prior to integrating it into our workflow. Obviously, we assume responsibility for all content.

%\bibliographystyle{apsrev4-2}
%\bibliography{references-v33}

\begin{thebibliography}{85}%
\makeatletter
\providecommand \@ifxundefined [1]{%
 \@ifx{#1\undefined}
}%
\providecommand \@ifnum [1]{%
 \ifnum #1\expandafter \@firstoftwo
 \else \expandafter \@secondoftwo
 \fi
}%
\providecommand \@ifx [1]{%
 \ifx #1\expandafter \@firstoftwo
 \else \expandafter \@secondoftwo
 \fi
}%
\providecommand \natexlab [1]{#1}%
\providecommand \enquote  [1]{``#1''}%
\providecommand \bibnamefont  [1]{#1}%
\providecommand \bibfnamefont [1]{#1}%
\providecommand \citenamefont [1]{#1}%
\providecommand \href@noop [0]{\@secondoftwo}%
\providecommand \href [0]{\begingroup \@sanitize@url \@href}%
\providecommand \@href[1]{\@@startlink{#1}\@@href}%
\providecommand \@@href[1]{\endgroup#1\@@endlink}%
\providecommand \@sanitize@url [0]{\catcode `\\12\catcode `\$12\catcode
  `\&12\catcode `\#12\catcode `\^12\catcode `\_12\catcode `\%12\relax}%
\providecommand \@@startlink[1]{}%
\providecommand \@@endlink[0]{}%
\providecommand \url  [0]{\begingroup\@sanitize@url \@url }%
\providecommand \@url [1]{\endgroup\@href {#1}{\urlprefix }}%
\providecommand \urlprefix  [0]{URL }%
\providecommand \Eprint [0]{\href }%
\providecommand \doibase [0]{https://doi.org/}%
\providecommand \selectlanguage [0]{\@gobble}%
\providecommand \bibinfo  [0]{\@secondoftwo}%
\providecommand \bibfield  [0]{\@secondoftwo}%
\providecommand \translation [1]{[#1]}%
\providecommand \BibitemOpen [0]{}%
\providecommand \bibitemStop [0]{}%
\providecommand \bibitemNoStop [0]{.\EOS\space}%
\providecommand \EOS [0]{\spacefactor3000\relax}%
\providecommand \BibitemShut  [1]{\csname bibitem#1\endcsname}%
\let\auto@bib@innerbib\@empty
%</preamble>
\bibitem [{\citenamefont {Newman}(2018)}]{Newman2018}%
  \BibitemOpen
  \bibfield  {author} {\bibinfo {author} {\bibfnamefont {M.}~\bibnamefont
  {Newman}},\ }\href
  {https://doi.org/https://doi.org/10.1093/oso/9780198805090.001.0001} {\emph
  {\bibinfo {title} {Networks}}},\ \bibinfo {edition} {2nd}\ ed.\ (\bibinfo
  {publisher} {Oxford University Press},\ \bibinfo {address} {Oxford, UK},\
  \bibinfo {year} {2018})\BibitemShut {NoStop}%
\bibitem [{\citenamefont {Barthelemy}(2022)}]{barthelemy2022}%
  \BibitemOpen
  \bibfield  {author} {\bibinfo {author} {\bibfnamefont {M.}~\bibnamefont
  {Barthelemy}},\ }\href
  {https://doi.org/https://doi.org/10.1007/978-3-030-94106-2} {\emph {\bibinfo
  {title} {Spatial Networks: A Complete Introduction: From Graph Theory and
  Statistical Physics to Real-World Applications}}}\ (\bibinfo  {publisher}
  {Springer},\ \bibinfo {address} {Cham, Switzerland},\ \bibinfo {year}
  {2022})\BibitemShut {NoStop}%
\bibitem [{\citenamefont {Ebrahimi-Nejad}\ \emph {et~al.}(2026)\citenamefont
  {Ebrahimi-Nejad}, \citenamefont {Rashidmanafi}, \citenamefont {Ravanbod},\
  and\ \citenamefont {Mollajafari}}]{ebra2026}%
  \BibitemOpen
  \bibfield  {author} {\bibinfo {author} {\bibfnamefont {S.}~\bibnamefont
  {Ebrahimi-Nejad}}, \bibinfo {author} {\bibfnamefont {A.}~\bibnamefont
  {Rashidmanafi}}, \bibinfo {author} {\bibfnamefont {S.}~\bibnamefont
  {Ravanbod}},\ and\ \bibinfo {author} {\bibfnamefont {M.}~\bibnamefont
  {Mollajafari}},\ }\href
  {https://doi.org/https://doi.org/10.1016/j.matdes.2026.116490} {\bibfield
  {journal} {\bibinfo  {journal} {Materials \& Design}\ }\textbf {\bibinfo
  {volume} {268}},\ \bibinfo {pages} {116490} (\bibinfo {year}
  {2026})}\BibitemShut {NoStop}%
\bibitem [{\citenamefont {Wytock}\ \emph {et~al.}(2026)\citenamefont {Wytock},
  \citenamefont {Daraio}, \citenamefont {Jaeger}, \citenamefont {Schuh},
  \citenamefont {Valdevit}, \citenamefont {Vitelli},\ and\ \citenamefont
  {Motter}}]{wytock2026}%
  \BibitemOpen
  \bibfield  {author} {\bibinfo {author} {\bibfnamefont {T.~P.}\ \bibnamefont
  {Wytock}}, \bibinfo {author} {\bibfnamefont {C.}~\bibnamefont {Daraio}},
  \bibinfo {author} {\bibfnamefont {H.~M.}\ \bibnamefont {Jaeger}}, \bibinfo
  {author} {\bibfnamefont {C.~A.}\ \bibnamefont {Schuh}}, \bibinfo {author}
  {\bibfnamefont {L.}~\bibnamefont {Valdevit}}, \bibinfo {author}
  {\bibfnamefont {V.}~\bibnamefont {Vitelli}},\ and\ \bibinfo {author}
  {\bibfnamefont {A.~E.}\ \bibnamefont {Motter}},\ }\href@noop {} {\bibfield
  {journal} {\bibinfo  {journal} {arXiv preprint}\ } (\bibinfo {year}
  {2026})},\ \bibinfo {note} {arXiv:2606.02695}\BibitemShut {NoStop}%
\bibitem [{\citenamefont {Obrero}\ \emph {et~al.}(2025)\citenamefont {Obrero},
  \citenamefont {Tirfe}, \citenamefont {Lee}, \citenamefont {Saptarshi},
  \citenamefont {Rock}, \citenamefont {Daniels},\ and\ \citenamefont
  {Newhall}}]{Obrero2024}%
  \BibitemOpen
  \bibfield  {author} {\bibinfo {author} {\bibfnamefont {C.}~\bibnamefont
  {Obrero}}, \bibinfo {author} {\bibfnamefont {M.}~\bibnamefont {Tirfe}},
  \bibinfo {author} {\bibfnamefont {C.}~\bibnamefont {Lee}}, \bibinfo {author}
  {\bibfnamefont {S.}~\bibnamefont {Saptarshi}}, \bibinfo {author}
  {\bibfnamefont {C.}~\bibnamefont {Rock}}, \bibinfo {author} {\bibfnamefont
  {K.~E.}\ \bibnamefont {Daniels}},\ and\ \bibinfo {author} {\bibfnamefont
  {K.~A.}\ \bibnamefont {Newhall}},\ }\href
  {https://doi.org/https://doi.org/10.1103/6bph-n6zj} {\bibfield  {journal}
  {\bibinfo  {journal} {Phys. Rev. E}\ }\textbf {\bibinfo {volume} {112}},\
  \bibinfo {pages} {035505} (\bibinfo {year} {2025})}\BibitemShut {NoStop}%
\bibitem [{\citenamefont {Liu}\ \emph {et~al.}(2025)\citenamefont {Liu},
  \citenamefont {Xia}, \citenamefont {Zhou},\ and\ \citenamefont
  {Wei}}]{Liu2025}%
  \BibitemOpen
  \bibfield  {author} {\bibinfo {author} {\bibfnamefont {Y.}~\bibnamefont
  {Liu}}, \bibinfo {author} {\bibfnamefont {B.}~\bibnamefont {Xia}}, \bibinfo
  {author} {\bibfnamefont {Y.}~\bibnamefont {Zhou}},\ and\ \bibinfo {author}
  {\bibfnamefont {K.}~\bibnamefont {Wei}},\ }\href
  {https://doi.org/10.1016/j.actamat.2024.120700} {\bibfield  {journal}
  {\bibinfo  {journal} {Acta Mater.}\ }\textbf {\bibinfo {volume} {285}},\
  \bibinfo {pages} {120700} (\bibinfo {year} {2025})}\BibitemShut {NoStop}%
\bibitem [{\citenamefont {Wang}\ \emph
  {et~al.}(2026{\natexlab{a}})\citenamefont {Wang}, \citenamefont {Maher},\
  and\ \citenamefont {Newhall}}]{wang_local_topology_2026}%
  \BibitemOpen
  \bibfield  {author} {\bibinfo {author} {\bibfnamefont {C.}~\bibnamefont
  {Wang}}, \bibinfo {author} {\bibfnamefont {C.~E.}\ \bibnamefont {Maher}},\
  and\ \bibinfo {author} {\bibfnamefont {K.~A.}\ \bibnamefont {Newhall}},\
  }\href {https://doi.org/10.1103/kqkn-mymm} {\bibfield  {journal} {\bibinfo
  {journal} {Phys. Rev. E}\ }\textbf {\bibinfo {volume} {114}},\ \bibinfo
  {pages} {024305} (\bibinfo {year} {2026}{\natexlab{a}})}\BibitemShut
  {NoStop}%
\bibitem [{\citenamefont {Barthelemy}\ and\ \citenamefont
  {Boeing}(2024)}]{Barthelemy2024}%
  \BibitemOpen
  \bibfield  {author} {\bibinfo {author} {\bibfnamefont {M.}~\bibnamefont
  {Barthelemy}}\ and\ \bibinfo {author} {\bibfnamefont {G.}~\bibnamefont
  {Boeing}},\ }\bibfield  {journal} {\bibinfo  {journal} {Findings}\ }\href
  {https://doi.org/10.32866/001c.122117} {10.32866/001c.122117} (\bibinfo
  {year} {2024})\BibitemShut {NoStop}%
\bibitem [{\citenamefont {Cuadra}\ \emph {et~al.}(2015)\citenamefont {Cuadra},
  \citenamefont {Salcedo-Sanz}, \citenamefont {Ser}, \citenamefont
  {Jim\'{e}nez-Fern\'{a}ndez},\ and\ \citenamefont {Geem}}]{Cuadra2015}%
  \BibitemOpen
  \bibfield  {author} {\bibinfo {author} {\bibfnamefont {L.}~\bibnamefont
  {Cuadra}}, \bibinfo {author} {\bibfnamefont {S.}~\bibnamefont
  {Salcedo-Sanz}}, \bibinfo {author} {\bibfnamefont {J.~D.}\ \bibnamefont
  {Ser}}, \bibinfo {author} {\bibfnamefont {S.}~\bibnamefont
  {Jim\'{e}nez-Fern\'{a}ndez}},\ and\ \bibinfo {author} {\bibfnamefont {Z.~W.}\
  \bibnamefont {Geem}},\ }\href {https://doi.org/10.3390/en8099211} {\bibfield
  {journal} {\bibinfo  {journal} {Energies}\ }\textbf {\bibinfo {volume} {8}},\
  \bibinfo {pages} {9211} (\bibinfo {year} {2015})}\BibitemShut {NoStop}%
\bibitem [{\citenamefont {Stiso}\ and\ \citenamefont
  {Bassett}(2018)}]{Stiso2018}%
  \BibitemOpen
  \bibfield  {author} {\bibinfo {author} {\bibfnamefont {J.}~\bibnamefont
  {Stiso}}\ and\ \bibinfo {author} {\bibfnamefont {D.~S.}\ \bibnamefont
  {Bassett}},\ }\href {https://doi.org/10.1016/j.tics.2018.09.007} {\bibfield
  {journal} {\bibinfo  {journal} {Trends Cogn. Sci.}\ }\textbf {\bibinfo
  {volume} {22}},\ \bibinfo {pages} {1127} (\bibinfo {year}
  {2018})}\BibitemShut {NoStop}%
\bibitem [{\citenamefont {Yang}\ \emph {et~al.}(2015)\citenamefont {Yang},
  \citenamefont {Wagner},\ and\ \citenamefont {Beli}}]{Yang2015}%
  \BibitemOpen
  \bibfield  {author} {\bibinfo {author} {\bibfnamefont {J.}~\bibnamefont
  {Yang}}, \bibinfo {author} {\bibfnamefont {S.~A.}\ \bibnamefont {Wagner}},\
  and\ \bibinfo {author} {\bibfnamefont {P.}~\bibnamefont {Beli}},\ }\href
  {https://doi.org/10.3389/fgene.2015.00344} {\bibfield  {journal} {\bibinfo
  {journal} {Front. Genet.}\ }\textbf {\bibinfo {volume} {6}},\ \bibinfo
  {pages} {344} (\bibinfo {year} {2015})}\BibitemShut {NoStop}%
\bibitem [{\citenamefont {Albert}\ \emph {et~al.}(2014)\citenamefont {Albert},
  \citenamefont {DasGupta},\ and\ \citenamefont {Mobasheri}}]{albert2014}%
  \BibitemOpen
  \bibfield  {author} {\bibinfo {author} {\bibfnamefont {R.}~\bibnamefont
  {Albert}}, \bibinfo {author} {\bibfnamefont {B.}~\bibnamefont {DasGupta}},\
  and\ \bibinfo {author} {\bibfnamefont {N.}~\bibnamefont {Mobasheri}},\ }\href
  {https://doi.org/10.1103/PhysRevE.89.032811} {\bibfield  {journal} {\bibinfo
  {journal} {Phys. Rev. E}\ }\textbf {\bibinfo {volume} {89}},\ \bibinfo
  {pages} {032811} (\bibinfo {year} {2014})}\BibitemShut {NoStop}%
\bibitem [{\citenamefont {Liu}\ and\ \citenamefont {Porter}(2020)}]{liu2020}%
  \BibitemOpen
  \bibfield  {author} {\bibinfo {author} {\bibfnamefont {A.}~\bibnamefont
  {Liu}}\ and\ \bibinfo {author} {\bibfnamefont {M.~A.}\ \bibnamefont
  {Porter}},\ }\href
  {https://doi.org/https://doi.org/10.1103/PhysRevE.101.062305} {\bibfield
  {journal} {\bibinfo  {journal} {Phys. Rev. E}\ }\textbf {\bibinfo {volume}
  {101}},\ \bibinfo {pages} {062305} (\bibinfo {year} {2020})}\BibitemShut
  {NoStop}%
\bibitem [{\citenamefont {Bogu$\tilde{\text{n}}$\'{a}}\ \emph
  {et~al.}(2021)\citenamefont {Bogu$\tilde{\text{n}}$\'{a}}, \citenamefont
  {Bonamassa}, \citenamefont {Domenico}, \citenamefont {Havlin}, \citenamefont
  {Krioukov},\ and\ \citenamefont {Serrano}}]{boguna_network_2021}%
  \BibitemOpen
  \bibfield  {author} {\bibinfo {author} {\bibfnamefont {M.}~\bibnamefont
  {Bogu$\tilde{\text{n}}$\'{a}}}, \bibinfo {author} {\bibfnamefont
  {I.}~\bibnamefont {Bonamassa}}, \bibinfo {author} {\bibfnamefont {M.~D.}\
  \bibnamefont {Domenico}}, \bibinfo {author} {\bibfnamefont {S.}~\bibnamefont
  {Havlin}}, \bibinfo {author} {\bibfnamefont {D.}~\bibnamefont {Krioukov}},\
  and\ \bibinfo {author} {\bibfnamefont {M.~A.}\ \bibnamefont {Serrano}},\
  }\href {https://doi.org/10.1038/s42254-020-00264-4} {\bibfield  {journal}
  {\bibinfo  {journal} {Nat. Rev. Phys.}\ }\textbf {\bibinfo {volume} {3}},\
  \bibinfo {pages} {114} (\bibinfo {year} {2021})}\BibitemShut {NoStop}%
\bibitem [{\citenamefont {Calka}(2009)}]{calka2009}%
  \BibitemOpen
  \bibfield  {author} {\bibinfo {author} {\bibfnamefont {P.}~\bibnamefont
  {Calka}},\ }in\ \href
  {https://doi.org/https://doi.org/10.1093/acprof:oso/9780199232574.001.0001}
  {\emph {\bibinfo {booktitle} {New Perspectives in Stochastic Geometry}}},\
  \bibinfo {editor} {edited by\ \bibinfo {editor} {\bibfnamefont {W.~S.}\
  \bibnamefont {Kendall}}\ and\ \bibinfo {editor} {\bibfnamefont
  {I.}~\bibnamefont {Molchanov}}}\ (\bibinfo  {publisher} {Oxford University
  Press},\ \bibinfo {address} {Oxford, UK},\ \bibinfo {year} {2009})\ pp.\
  \bibinfo {pages} {145--170}\BibitemShut {NoStop}%
\bibitem [{\citenamefont {Yadav}\ and\ \citenamefont {Xia}(2025)}]{yadav2025}%
  \BibitemOpen
  \bibfield  {author} {\bibinfo {author} {\bibfnamefont {Y.}~\bibnamefont
  {Yadav}}\ and\ \bibinfo {author} {\bibfnamefont {K.}~\bibnamefont {Xia}},\
  }\href@noop {} {\bibfield  {journal} {\bibinfo  {journal} {arXiv preprint}\ }
  (\bibinfo {year} {2025})},\ \bibinfo {note} {arXiv:2510.22599}\BibitemShut
  {NoStop}%
\bibitem [{\citenamefont {Devriendt}\ and\ \citenamefont
  {Lambiotte}(2022)}]{dev2022}%
  \BibitemOpen
  \bibfield  {author} {\bibinfo {author} {\bibfnamefont {K.}~\bibnamefont
  {Devriendt}}\ and\ \bibinfo {author} {\bibfnamefont {R.}~\bibnamefont
  {Lambiotte}},\ }\href {https://doi.org/10.1088/2632-072X/ac730d} {\bibfield
  {journal} {\bibinfo  {journal} {J. Phys. Complex.}\ }\textbf {\bibinfo
  {volume} {3}},\ \bibinfo {pages} {025008} (\bibinfo {year}
  {2022})}\BibitemShut {NoStop}%
\bibitem [{\citenamefont {Devriendt}\ \emph {et~al.}(2024)\citenamefont
  {Devriendt}, \citenamefont {Ottolini},\ and\ \citenamefont
  {Steinerberger}}]{dev2024}%
  \BibitemOpen
  \bibfield  {author} {\bibinfo {author} {\bibfnamefont {K.}~\bibnamefont
  {Devriendt}}, \bibinfo {author} {\bibfnamefont {A.}~\bibnamefont
  {Ottolini}},\ and\ \bibinfo {author} {\bibfnamefont {S.}~\bibnamefont
  {Steinerberger}},\ }\href
  {https://doi.org/https://doi.org/10.1016/j.dam.2024.01.012} {\bibfield
  {journal} {\bibinfo  {journal} {Disc. App. Math.}\ }\textbf {\bibinfo
  {volume} {348}},\ \bibinfo {pages} {68} (\bibinfo {year} {2024})}\BibitemShut
  {NoStop}%
\bibitem [{\citenamefont {Robertson}\ \emph {et~al.}(2024)\citenamefont
  {Robertson}, \citenamefont {Wan},\ and\ \citenamefont
  {Cloninger}}]{robertson2024}%
  \BibitemOpen
  \bibfield  {author} {\bibinfo {author} {\bibfnamefont {S.}~\bibnamefont
  {Robertson}}, \bibinfo {author} {\bibfnamefont {Z.}~\bibnamefont {Wan}},\
  and\ \bibinfo {author} {\bibfnamefont {A.}~\bibnamefont {Cloninger}},\
  }\href@noop {} {\bibfield  {journal} {\bibinfo  {journal} {arXiv preprint}\ }
  (\bibinfo {year} {2024})},\ \bibinfo {note} {arXiv:2404.15261}\BibitemShut
  {NoStop}%
\bibitem [{\citenamefont {Ollivier}(2009)}]{ollivier2009}%
  \BibitemOpen
  \bibfield  {author} {\bibinfo {author} {\bibfnamefont {Y.}~\bibnamefont
  {Ollivier}},\ }\href
  {https://doi.org/https://doi.org/10.1016/j.jfa.2008.11.001} {\bibfield
  {journal} {\bibinfo  {journal} {J. Funct. Anal.}\ }\textbf {\bibinfo {volume}
  {256}},\ \bibinfo {pages} {810} (\bibinfo {year} {2009})}\BibitemShut
  {NoStop}%
\bibitem [{\citenamefont {Ollivier}(2010)}]{Ollivier2010}%
  \BibitemOpen
  \bibfield  {author} {\bibinfo {author} {\bibfnamefont {Y.}~\bibnamefont
  {Ollivier}},\ }in\ \href {https://doi.org/https://doi.org/10.1142/e025}
  {\emph {\bibinfo {booktitle} {Probabilistic Approach to Geometry}}},\
  Vol.~\bibinfo {volume} {57},\ \bibinfo {editor} {edited by\ \bibinfo {editor}
  {\bibfnamefont {M.}~\bibnamefont {Kotani}}, \bibinfo {editor} {\bibfnamefont
  {M.}~\bibnamefont {Hino}},\ and\ \bibinfo {editor} {\bibfnamefont
  {T.}~\bibnamefont {Kumagai}}}\ (\bibinfo  {publisher} {Mathematical Society
  of Japan},\ \bibinfo {address} {Tokyo, Japan},\ \bibinfo {year} {2010})\ pp.\
  \bibinfo {pages} {343--381}\BibitemShut {NoStop}%
\bibitem [{\citenamefont {Sreejith}\ \emph {et~al.}(2016)\citenamefont
  {Sreejith}, \citenamefont {Mohanraj}, \citenamefont {Jost}, \citenamefont
  {Saucan},\ and\ \citenamefont {Samal}}]{Sreejith_2016}%
  \BibitemOpen
  \bibfield  {author} {\bibinfo {author} {\bibfnamefont {R.~P.}\ \bibnamefont
  {Sreejith}}, \bibinfo {author} {\bibfnamefont {K.}~\bibnamefont {Mohanraj}},
  \bibinfo {author} {\bibfnamefont {J.}~\bibnamefont {Jost}}, \bibinfo {author}
  {\bibfnamefont {E.}~\bibnamefont {Saucan}},\ and\ \bibinfo {author}
  {\bibfnamefont {A.}~\bibnamefont {Samal}},\ }\href
  {https://doi.org/10.1088/1742-5468/2016/06/063206} {\bibfield  {journal}
  {\bibinfo  {journal} {J. Stat. Mech. Theory Exp.}\ }\textbf {\bibinfo
  {volume} {2016}},\ \bibinfo {pages} {063206} (\bibinfo {year}
  {2016})}\BibitemShut {NoStop}%
\bibitem [{\citenamefont {Villani}(2009)}]{Villani2009}%
  \BibitemOpen
  \bibfield  {author} {\bibinfo {author} {\bibfnamefont {C.}~\bibnamefont
  {Villani}},\ }\href {https://doi.org/10.1007/978-3-540-71050-9} {\emph
  {\bibinfo {title} {Optimal Transport: Old and New}}}\ (\bibinfo  {publisher}
  {Springer},\ \bibinfo {address} {Heidelberg, Germany},\ \bibinfo {year}
  {2009})\BibitemShut {NoStop}%
\bibitem [{\citenamefont {Sawchuk}\ and\ \citenamefont
  {Sivak}(2026)}]{sawchuk2026}%
  \BibitemOpen
  \bibfield  {author} {\bibinfo {author} {\bibfnamefont {J.~R.}\ \bibnamefont
  {Sawchuk}}\ and\ \bibinfo {author} {\bibfnamefont {D.~A.}\ \bibnamefont
  {Sivak}},\ }\href@noop {} {\bibfield  {journal} {\bibinfo  {journal} {arXiv
  preprint}\ } (\bibinfo {year} {2026})},\ \bibinfo {note}
  {arXiv:2601.01273}\BibitemShut {NoStop}%
\bibitem [{\citenamefont {Torquato}\ and\ \citenamefont
  {Stillinger}(2003)}]{torquato_local_2003}%
  \BibitemOpen
  \bibfield  {author} {\bibinfo {author} {\bibfnamefont {S.}~\bibnamefont
  {Torquato}}\ and\ \bibinfo {author} {\bibfnamefont {F.~H.}\ \bibnamefont
  {Stillinger}},\ }\href {https://doi.org/10.1103/PhysRevE.68.041113}
  {\bibfield  {journal} {\bibinfo  {journal} {Phys. Rev. E}\ }\textbf {\bibinfo
  {volume} {68}},\ \bibinfo {pages} {041113} (\bibinfo {year}
  {2003})}\BibitemShut {NoStop}%
\bibitem [{\citenamefont {Torquato}(2018)}]{torquato_hyperuniform_2018}%
  \BibitemOpen
  \bibfield  {author} {\bibinfo {author} {\bibfnamefont {S.}~\bibnamefont
  {Torquato}},\ }\href {https://doi.org/10.1016/j.physrep.2018.03.001}
  {\bibfield  {journal} {\bibinfo  {journal} {Phys. Rep.}\ }\textbf {\bibinfo
  {volume} {745}},\ \bibinfo {pages} {1} (\bibinfo {year} {2018})}\BibitemShut
  {NoStop}%
\bibitem [{\citenamefont {Martin}\ and\ \citenamefont
  {Yalcin}(1980)}]{Martin1980}%
  \BibitemOpen
  \bibfield  {author} {\bibinfo {author} {\bibfnamefont {P.~A.}\ \bibnamefont
  {Martin}}\ and\ \bibinfo {author} {\bibfnamefont {T.}~\bibnamefont
  {Yalcin}},\ }\href {https://doi.org/10.1007/BF01012866} {\bibfield  {journal}
  {\bibinfo  {journal} {J. Stat. Phys.}\ }\textbf {\bibinfo {volume} {22}},\
  \bibinfo {pages} {435} (\bibinfo {year} {1980})}\BibitemShut {NoStop}%
\bibitem [{\citenamefont {Lebowitz}(1983)}]{Lebowitz1983}%
  \BibitemOpen
  \bibfield  {author} {\bibinfo {author} {\bibfnamefont {J.~L.}\ \bibnamefont
  {Lebowitz}},\ }\href {https://doi.org/10.1103/PhysRevA.27.1491} {\bibfield
  {journal} {\bibinfo  {journal} {Phys. Rev. A}\ }\textbf {\bibinfo {volume}
  {27}},\ \bibinfo {pages} {1491} (\bibinfo {year} {1983})}\BibitemShut
  {NoStop}%
\bibitem [{\citenamefont {Martin}(1988)}]{Martin1988}%
  \BibitemOpen
  \bibfield  {author} {\bibinfo {author} {\bibfnamefont {P.~A.}\ \bibnamefont
  {Martin}},\ }\href {https://doi.org/10.1103/RevModPhys.60.1075} {\bibfield
  {journal} {\bibinfo  {journal} {Rev. Mod. Phys.}\ }\textbf {\bibinfo {volume}
  {60}},\ \bibinfo {pages} {1075} (\bibinfo {year} {1988})}\BibitemShut
  {NoStop}%
\bibitem [{\citenamefont {Jancovici}\ \emph {et~al.}(1993)\citenamefont
  {Jancovici}, \citenamefont {Lebowitz},\ and\ \citenamefont
  {Manificat}}]{Jancovici1993}%
  \BibitemOpen
  \bibfield  {author} {\bibinfo {author} {\bibfnamefont {B.}~\bibnamefont
  {Jancovici}}, \bibinfo {author} {\bibfnamefont {J.~L.}\ \bibnamefont
  {Lebowitz}},\ and\ \bibinfo {author} {\bibfnamefont {G.}~\bibnamefont
  {Manificat}},\ }\href {https://doi.org/10.1007/BF01048032} {\bibfield
  {journal} {\bibinfo  {journal} {J. Stat. Phys.}\ }\textbf {\bibinfo {volume}
  {72}},\ \bibinfo {pages} {773} (\bibinfo {year} {1993})}\BibitemShut
  {NoStop}%
\bibitem [{\citenamefont {Levesque}\ \emph {et~al.}(2000)\citenamefont
  {Levesque}, \citenamefont {Weis},\ and\ \citenamefont
  {Lebowitz}}]{Levesque2000}%
  \BibitemOpen
  \bibfield  {author} {\bibinfo {author} {\bibfnamefont {D.}~\bibnamefont
  {Levesque}}, \bibinfo {author} {\bibfnamefont {J.-J.}\ \bibnamefont {Weis}},\
  and\ \bibinfo {author} {\bibfnamefont {J.~L.}\ \bibnamefont {Lebowitz}},\
  }\href {https://doi.org/10.1023/A:1018643829340} {\bibfield  {journal}
  {\bibinfo  {journal} {J. Stat. Phys.}\ }\textbf {\bibinfo {volume} {100}},\
  \bibinfo {pages} {209} (\bibinfo {year} {2000})}\BibitemShut {NoStop}%
\bibitem [{\citenamefont {Lachi\`{e}ze-Rey}(2025)}]{lach2025}%
  \BibitemOpen
  \bibfield  {author} {\bibinfo {author} {\bibfnamefont {R.}~\bibnamefont
  {Lachi\`{e}ze-Rey}},\ }\href@noop {} {\bibfield  {journal} {\bibinfo
  {journal} {arXiv preprint}\ } (\bibinfo {year} {2025})},\ \bibinfo {note}
  {arXiv:2510.18392}\BibitemShut {NoStop}%
\bibitem [{\citenamefont {Torquato}\ and\ \citenamefont
  {Chen}(2018)}]{torquato_multifunctional_2018}%
  \BibitemOpen
  \bibfield  {author} {\bibinfo {author} {\bibfnamefont {S.}~\bibnamefont
  {Torquato}}\ and\ \bibinfo {author} {\bibfnamefont {D.}~\bibnamefont
  {Chen}},\ }\href {https://doi.org/10.1088/2399-7532/aaca91} {\bibfield
  {journal} {\bibinfo  {journal} {Multifunct. Mater.}\ }\textbf {\bibinfo
  {volume} {1}},\ \bibinfo {pages} {015001} (\bibinfo {year}
  {2018})}\BibitemShut {NoStop}%
\bibitem [{\citenamefont {Florescu}\ \emph {et~al.}(2009)\citenamefont
  {Florescu}, \citenamefont {Torquato},\ and\ \citenamefont
  {Steinhardt}}]{Florescu2009}%
  \BibitemOpen
  \bibfield  {author} {\bibinfo {author} {\bibfnamefont {M.}~\bibnamefont
  {Florescu}}, \bibinfo {author} {\bibfnamefont {S.}~\bibnamefont {Torquato}},\
  and\ \bibinfo {author} {\bibfnamefont {P.~J.}\ \bibnamefont {Steinhardt}},\
  }\href {https://doi.org/10.1073/pnas.0907744106} {\bibfield  {journal}
  {\bibinfo  {journal} {Proc. Natl. Acad. Sci. U.S.A.}\ }\textbf {\bibinfo
  {volume} {106}},\ \bibinfo {pages} {20658} (\bibinfo {year}
  {2009})}\BibitemShut {NoStop}%
\bibitem [{\citenamefont {Man}\ \emph {et~al.}(2013)\citenamefont {Man},
  \citenamefont {Florescu}, \citenamefont {Williamson}, \citenamefont {He},
  \citenamefont {Hashemizad}, \citenamefont {Leung}, \citenamefont {Liner},
  \citenamefont {Torquato}, \citenamefont {Chaikin},\ and\ \citenamefont
  {Steinhardt}}]{Man2013}%
  \BibitemOpen
  \bibfield  {author} {\bibinfo {author} {\bibfnamefont {W.}~\bibnamefont
  {Man}}, \bibinfo {author} {\bibfnamefont {M.}~\bibnamefont {Florescu}},
  \bibinfo {author} {\bibfnamefont {E.~P.}\ \bibnamefont {Williamson}},
  \bibinfo {author} {\bibfnamefont {Y.}~\bibnamefont {He}}, \bibinfo {author}
  {\bibfnamefont {S.~R.}\ \bibnamefont {Hashemizad}}, \bibinfo {author}
  {\bibfnamefont {B.~Y.~C.}\ \bibnamefont {Leung}}, \bibinfo {author}
  {\bibfnamefont {D.~R.}\ \bibnamefont {Liner}}, \bibinfo {author}
  {\bibfnamefont {S.}~\bibnamefont {Torquato}}, \bibinfo {author}
  {\bibfnamefont {P.~M.}\ \bibnamefont {Chaikin}},\ and\ \bibinfo {author}
  {\bibfnamefont {P.~J.}\ \bibnamefont {Steinhardt}},\ }\href
  {https://doi.org/10.1073/pnas.1307879110} {\bibfield  {journal} {\bibinfo
  {journal} {Proc. Natl. Acad. Sci. U.S.A.}\ }\textbf {\bibinfo {volume}
  {110}},\ \bibinfo {pages} {15886} (\bibinfo {year} {2013})}\BibitemShut
  {NoStop}%
\bibitem [{\citenamefont {Klatt}\ \emph {et~al.}(2022)\citenamefont {Klatt},
  \citenamefont {Steinhardt},\ and\ \citenamefont {Torquato}}]{Klatt2022}%
  \BibitemOpen
  \bibfield  {author} {\bibinfo {author} {\bibfnamefont {M.~A.}\ \bibnamefont
  {Klatt}}, \bibinfo {author} {\bibfnamefont {P.~J.}\ \bibnamefont
  {Steinhardt}},\ and\ \bibinfo {author} {\bibfnamefont {S.}~\bibnamefont
  {Torquato}},\ }\href {https://doi.org/10.1073/pnas.2213633119} {\bibfield
  {journal} {\bibinfo  {journal} {Proc. Natl. Acad. Sci.}\ }\textbf {\bibinfo
  {volume} {119}},\ \bibinfo {pages} {e2213633119} (\bibinfo {year}
  {2022})}\BibitemShut {NoStop}%
\bibitem [{\citenamefont {Siedentop}\ \emph {et~al.}(2024)\citenamefont
  {Siedentop}, \citenamefont {Lui}, \citenamefont {Maret}, \citenamefont
  {Chaikin}, \citenamefont {Steinhardt}, \citenamefont {Torquato},
  \citenamefont {Keim},\ and\ \citenamefont {Florescu}}]{Siedentop2024}%
  \BibitemOpen
  \bibfield  {author} {\bibinfo {author} {\bibfnamefont {L.}~\bibnamefont
  {Siedentop}}, \bibinfo {author} {\bibfnamefont {G.}~\bibnamefont {Lui}},
  \bibinfo {author} {\bibfnamefont {G.}~\bibnamefont {Maret}}, \bibinfo
  {author} {\bibfnamefont {P.~M.}\ \bibnamefont {Chaikin}}, \bibinfo {author}
  {\bibfnamefont {P.~J.}\ \bibnamefont {Steinhardt}}, \bibinfo {author}
  {\bibfnamefont {S.}~\bibnamefont {Torquato}}, \bibinfo {author}
  {\bibfnamefont {P.}~\bibnamefont {Keim}},\ and\ \bibinfo {author}
  {\bibfnamefont {M.}~\bibnamefont {Florescu}},\ }\href
  {https://doi.org/10.1093/pnasnexus/pgae383} {\bibfield  {journal} {\bibinfo
  {journal} {PNAS Nexus}\ }\textbf {\bibinfo {volume} {3}},\ \bibinfo {pages}
  {pgae383} (\bibinfo {year} {2024})}\BibitemShut {NoStop}%
\bibitem [{\citenamefont {Moody}\ \emph
  {et~al.}(2026{\natexlab{a}})\citenamefont {Moody}, \citenamefont {Maher},
  \citenamefont {Gwalani},\ and\ \citenamefont {Rock}}]{moody2026alloy}%
  \BibitemOpen
  \bibfield  {author} {\bibinfo {author} {\bibfnamefont {K.}~\bibnamefont
  {Moody}}, \bibinfo {author} {\bibfnamefont {C.~E.}\ \bibnamefont {Maher}},
  \bibinfo {author} {\bibfnamefont {B.}~\bibnamefont {Gwalani}},\ and\ \bibinfo
  {author} {\bibfnamefont {C.}~\bibnamefont {Rock}},\ }in\ \href
  {https://doi.org/10.1007/978-3-032-35176-0_16} {\emph {\bibinfo {booktitle}
  {Proceedings of the 11th International Symposium on Superalloy 718 and
  Derivatives}}},\ \bibinfo {series and number} {The Minerals, Metals \&
  Materials Series},\ \bibinfo {editor} {edited by\ \bibinfo {editor}
  {\bibfnamefont {E.~A.}\ \bibnamefont {Ott}}, \bibinfo {editor} {\bibfnamefont
  {C.}~\bibnamefont {Sudbrack}}, \bibinfo {editor} {\bibfnamefont
  {A.}~\bibnamefont {Wessman}}, \bibinfo {editor} {\bibfnamefont
  {J.}~\bibnamefont {Andersson}}, \bibinfo {editor} {\bibfnamefont
  {Z.}~\bibnamefont {Bi}}, \bibinfo {editor} {\bibfnamefont {I.}~\bibnamefont
  {Dempster}}, \bibinfo {editor} {\bibfnamefont {M.}~\bibnamefont {Detrois}},
  \bibinfo {editor} {\bibfnamefont {M.}~\bibnamefont {Fahrmann}}, \bibinfo
  {editor} {\bibfnamefont {M.}~\bibnamefont {Kirka}}, \bibinfo {editor}
  {\bibfnamefont {J.}~\bibnamefont {Klemm-Toole}}, \bibinfo {editor}
  {\bibfnamefont {D.}~\bibnamefont {Nagahama}}, \bibinfo {editor}
  {\bibfnamefont {J.}~\bibnamefont {Rame}}, \bibinfo {editor} {\bibfnamefont
  {K.}~\bibnamefont {Severs}},\ and\ \bibinfo {editor} {\bibfnamefont
  {K.}~\bibnamefont {Unocic}}}\ (\bibinfo  {publisher} {Springer},\ \bibinfo
  {address} {Cham, Switzerland},\ \bibinfo {year} {2026})\ pp.\ \bibinfo
  {pages} {295--308}\BibitemShut {NoStop}%
\bibitem [{\citenamefont {Maher}\ and\ \citenamefont
  {Newhall}(2025)}]{maher_newhall_network_hyperuniform}%
  \BibitemOpen
  \bibfield  {author} {\bibinfo {author} {\bibfnamefont {C.~E.}\ \bibnamefont
  {Maher}}\ and\ \bibinfo {author} {\bibfnamefont {K.~A.}\ \bibnamefont
  {Newhall}},\ }\href {https://doi.org/https://doi.org/10.1103/l5hk-kwyn}
  {\bibfield  {journal} {\bibinfo  {journal} {Phys. Rev. E}\ }\textbf {\bibinfo
  {volume} {111}},\ \bibinfo {pages} {065420} (\bibinfo {year}
  {2025})}\BibitemShut {NoStop}%
\bibitem [{\citenamefont {Newby}\ \emph
  {et~al.}(2025{\natexlab{a}})\citenamefont {Newby}, \citenamefont {Shi},
  \citenamefont {Jiao}, \citenamefont {Albert},\ and\ \citenamefont
  {Torquato}}]{Newby2025a}%
  \BibitemOpen
  \bibfield  {author} {\bibinfo {author} {\bibfnamefont {E.}~\bibnamefont
  {Newby}}, \bibinfo {author} {\bibfnamefont {W.}~\bibnamefont {Shi}}, \bibinfo
  {author} {\bibfnamefont {Y.}~\bibnamefont {Jiao}}, \bibinfo {author}
  {\bibfnamefont {R.}~\bibnamefont {Albert}},\ and\ \bibinfo {author}
  {\bibfnamefont {S.}~\bibnamefont {Torquato}},\ }\href
  {https://doi.org/10.1103/PhysRevE.111.034123} {\bibfield  {journal} {\bibinfo
   {journal} {Phys. Rev. E}\ }\textbf {\bibinfo {volume} {111}},\ \bibinfo
  {pages} {034123} (\bibinfo {year} {2025}{\natexlab{a}})}\BibitemShut
  {NoStop}%
\bibitem [{\citenamefont {Newby}\ \emph
  {et~al.}(2025{\natexlab{b}})\citenamefont {Newby}, \citenamefont {Shi},
  \citenamefont {Jiao}, \citenamefont {Torquato},\ and\ \citenamefont
  {Albert}}]{Newby2025b}%
  \BibitemOpen
  \bibfield  {author} {\bibinfo {author} {\bibfnamefont {E.}~\bibnamefont
  {Newby}}, \bibinfo {author} {\bibfnamefont {W.}~\bibnamefont {Shi}}, \bibinfo
  {author} {\bibfnamefont {Y.}~\bibnamefont {Jiao}}, \bibinfo {author}
  {\bibfnamefont {S.}~\bibnamefont {Torquato}},\ and\ \bibinfo {author}
  {\bibfnamefont {R.}~\bibnamefont {Albert}},\ }\href
  {https://doi.org/10.1093/comnet/cnaf040} {\bibfield  {journal} {\bibinfo
  {journal} {J. Cplx. Netw.}\ }\textbf {\bibinfo {volume} {13}},\ \bibinfo
  {pages} {cnaf040} (\bibinfo {year} {2025}{\natexlab{b}})}\BibitemShut
  {NoStop}%
\bibitem [{\citenamefont {Wang}\ \emph
  {et~al.}(2026{\natexlab{b}})\citenamefont {Wang}, \citenamefont {Kim},
  \citenamefont {Jiao}, \citenamefont {Stuhl}, \citenamefont {Torquato},\ and\
  \citenamefont {Albert}}]{wang2026}%
  \BibitemOpen
  \bibfield  {author} {\bibinfo {author} {\bibfnamefont {Y.}~\bibnamefont
  {Wang}}, \bibinfo {author} {\bibfnamefont {J.}~\bibnamefont {Kim}}, \bibinfo
  {author} {\bibfnamefont {Y.}~\bibnamefont {Jiao}}, \bibinfo {author}
  {\bibfnamefont {I.}~\bibnamefont {Stuhl}}, \bibinfo {author} {\bibfnamefont
  {S.}~\bibnamefont {Torquato}},\ and\ \bibinfo {author} {\bibfnamefont
  {R.}~\bibnamefont {Albert}},\ }\href@noop {} {\bibfield  {journal} {\bibinfo
  {journal} {arXiv preprint}\ } (\bibinfo {year} {2026}{\natexlab{b}})},\
  \bibinfo {note} {arXiv:2603.16167}\BibitemShut {NoStop}%
\bibitem [{\citenamefont {Maher}\ \emph {et~al.}(2026)\citenamefont {Maher},
  \citenamefont {Marzuola},\ and\ \citenamefont {Newhall}}]{maher2026spectrum}%
  \BibitemOpen
  \bibfield  {author} {\bibinfo {author} {\bibfnamefont {C.~E.}\ \bibnamefont
  {Maher}}, \bibinfo {author} {\bibfnamefont {J.~L.}\ \bibnamefont
  {Marzuola}},\ and\ \bibinfo {author} {\bibfnamefont {K.~A.}\ \bibnamefont
  {Newhall}},\ }\href@noop {} {\bibfield  {journal} {\bibinfo  {journal} {arXiv
  preprint}\ } (\bibinfo {year} {2026})},\ \bibinfo {note}
  {arXiv:2602.22341}\BibitemShut {NoStop}%
\bibitem [{\citenamefont {Gabrielli}(2004)}]{Gabrielli2004}%
  \BibitemOpen
  \bibfield  {author} {\bibinfo {author} {\bibfnamefont {A.}~\bibnamefont
  {Gabrielli}},\ }\href {https://doi.org/10.1103/PhysRevE.70.066131} {\bibfield
   {journal} {\bibinfo  {journal} {Phys. Rev. E}\ }\textbf {\bibinfo {volume}
  {70}},\ \bibinfo {pages} {066131} (\bibinfo {year} {2004})}\BibitemShut
  {NoStop}%
\bibitem [{\citenamefont {Vanoni}\ \emph {et~al.}(2025)\citenamefont {Vanoni},
  \citenamefont {Steinhardt},\ and\ \citenamefont {Torquato}}]{Vanoni2025}%
  \BibitemOpen
  \bibfield  {author} {\bibinfo {author} {\bibfnamefont {C.}~\bibnamefont
  {Vanoni}}, \bibinfo {author} {\bibfnamefont {P.~J.}\ \bibnamefont
  {Steinhardt}},\ and\ \bibinfo {author} {\bibfnamefont {S.}~\bibnamefont
  {Torquato}},\ }\href {https://doi.org/10.1103/511q-ltf1} {\bibfield
  {journal} {\bibinfo  {journal} {Phys. Rev. E}\ }\textbf {\bibinfo {volume}
  {112}},\ \bibinfo {pages} {044142} (\bibinfo {year} {2025})}\BibitemShut
  {NoStop}%
\bibitem [{\citenamefont {Torquato}(2016)}]{Torquato2016}%
  \BibitemOpen
  \bibfield  {author} {\bibinfo {author} {\bibfnamefont {S.}~\bibnamefont
  {Torquato}},\ }\href {https://doi.org/10.1088/0953-8984/28/41/414012}
  {\bibfield  {journal} {\bibinfo  {journal} {J. Phys. Cond. Mat.}\ }\textbf
  {\bibinfo {volume} {28}},\ \bibinfo {pages} {414012} (\bibinfo {year}
  {2016})}\BibitemShut {NoStop}%
\bibitem [{\citenamefont {Kim}\ and\ \citenamefont {Torquato}(2021)}]{kim2021}%
  \BibitemOpen
  \bibfield  {author} {\bibinfo {author} {\bibfnamefont {J.}~\bibnamefont
  {Kim}}\ and\ \bibinfo {author} {\bibfnamefont {S.}~\bibnamefont {Torquato}},\
  }\href {https://doi.org/https://doi.org/10.1103/PhysRevE.103.012123}
  {\bibfield  {journal} {\bibinfo  {journal} {Phys. Rev. E}\ }\textbf {\bibinfo
  {volume} {103}},\ \bibinfo {pages} {012123} (\bibinfo {year}
  {2021})}\BibitemShut {NoStop}%
\bibitem [{\citenamefont {Azarhooshang}\ \emph {et~al.}(2020)\citenamefont
  {Azarhooshang}, \citenamefont {Sengupta},\ and\ \citenamefont
  {DasGupta}}]{Azarhooshang2020}%
  \BibitemOpen
  \bibfield  {author} {\bibinfo {author} {\bibfnamefont {N.}~\bibnamefont
  {Azarhooshang}}, \bibinfo {author} {\bibfnamefont {P.}~\bibnamefont
  {Sengupta}},\ and\ \bibinfo {author} {\bibfnamefont {B.}~\bibnamefont
  {DasGupta}},\ }\href {https://doi.org/10.3390/math8091416} {\bibfield
  {journal} {\bibinfo  {journal} {Mathematics}\ }\textbf {\bibinfo {volume}
  {8}},\ \bibinfo {pages} {1416} (\bibinfo {year} {2020})}\BibitemShut
  {NoStop}%
\bibitem [{\citenamefont {{do Carmo}}(1992)}]{docarmo1992}%
  \BibitemOpen
  \bibfield  {author} {\bibinfo {author} {\bibfnamefont {M.~P.}\ \bibnamefont
  {{do Carmo}}},\ }\href@noop {} {\emph {\bibinfo {title} {Riemannian
  Geometry}}}\ (\bibinfo  {publisher} {Birkh\"{a}user},\ \bibinfo {address}
  {Basel, Switzerland},\ \bibinfo {year} {1992})\BibitemShut {NoStop}%
\bibitem [{\citenamefont {Jost}\ and\ \citenamefont {Liu}(2014)}]{jost2014}%
  \BibitemOpen
  \bibfield  {author} {\bibinfo {author} {\bibfnamefont {J.}~\bibnamefont
  {Jost}}\ and\ \bibinfo {author} {\bibfnamefont {S.}~\bibnamefont {Liu}},\
  }\href {https://doi.org/https://doi.org/10.1007/s00454-013-9558-1} {\bibfield
   {journal} {\bibinfo  {journal} {Discrete Comput. Geom.}\ }\textbf {\bibinfo
  {volume} {51}},\ \bibinfo {pages} {300} (\bibinfo {year} {2014})}\BibitemShut
  {NoStop}%
\bibitem [{\citenamefont {Ellens}\ \emph {et~al.}(2011)\citenamefont {Ellens},
  \citenamefont {Spieksma}, \citenamefont {Mieghem}, \citenamefont
  {Jamakovic},\ and\ \citenamefont {Kooij}}]{Ellens2011}%
  \BibitemOpen
  \bibfield  {author} {\bibinfo {author} {\bibfnamefont {W.}~\bibnamefont
  {Ellens}}, \bibinfo {author} {\bibfnamefont {F.~M.}\ \bibnamefont
  {Spieksma}}, \bibinfo {author} {\bibfnamefont {P.~V.}\ \bibnamefont
  {Mieghem}}, \bibinfo {author} {\bibfnamefont {A.}~\bibnamefont {Jamakovic}},\
  and\ \bibinfo {author} {\bibfnamefont {R.~E.}\ \bibnamefont {Kooij}},\ }\href
  {https://doi.org/10.1016/j.laa.2011.02.024} {\bibfield  {journal} {\bibinfo
  {journal} {Lin. Alg. Appl.}\ }\textbf {\bibinfo {volume} {435}},\ \bibinfo
  {pages} {2491} (\bibinfo {year} {2011})}\BibitemShut {NoStop}%
\bibitem [{\citenamefont {Montenegro}\ and\ \citenamefont
  {Tetali}(2006)}]{montenegro2006}%
  \BibitemOpen
  \bibfield  {author} {\bibinfo {author} {\bibfnamefont {R.}~\bibnamefont
  {Montenegro}}\ and\ \bibinfo {author} {\bibfnamefont {P.}~\bibnamefont
  {Tetali}},\ }\href {https://doi.org/https://doi.org/10.1561/0400000003}
  {\bibfield  {journal} {\bibinfo  {journal} {Found. Trend. Theor. Comp. Sci.}\
  }\textbf {\bibinfo {volume} {1}},\ \bibinfo {pages} {237} (\bibinfo {year}
  {2006})}\BibitemShut {NoStop}%
\bibitem [{\citenamefont {Raj}\ \emph {et~al.}(2026)\citenamefont {Raj},
  \citenamefont {Sun}, \citenamefont {Maher}, \citenamefont {Newhall},\ and\
  \citenamefont {Porter}}]{Raj2026HuPPICode}%
  \BibitemOpen
  \bibfield  {author} {\bibinfo {author} {\bibfnamefont {J.~V.}\ \bibnamefont
  {Raj}}, \bibinfo {author} {\bibfnamefont {X.}~\bibnamefont {Sun}}, \bibinfo
  {author} {\bibfnamefont {C.~E.}\ \bibnamefont {Maher}}, \bibinfo {author}
  {\bibfnamefont {K.~A.}\ \bibnamefont {Newhall}},\ and\ \bibinfo {author}
  {\bibfnamefont {M.~A.}\ \bibnamefont {Porter}},\ }\href
  {https://doi.org/10.5281/zenodo.23274708} {\bibinfo {title} {{HuPPI} network
  analysis}},\ \bibinfo {howpublished} {Zenodo} (\bibinfo {year} {2026}),\
  \bibinfo {note} {version v1.0.0}\BibitemShut {NoStop}%
\bibitem [{\citenamefont {Gabrielli}\ \emph {et~al.}(2002)\citenamefont
  {Gabrielli}, \citenamefont {Joyce},\ and\ \citenamefont
  {Labini}}]{gabrielli2002}%
  \BibitemOpen
  \bibfield  {author} {\bibinfo {author} {\bibfnamefont {A.}~\bibnamefont
  {Gabrielli}}, \bibinfo {author} {\bibfnamefont {M.}~\bibnamefont {Joyce}},\
  and\ \bibinfo {author} {\bibfnamefont {F.~S.}\ \bibnamefont {Labini}},\
  }\href {https://doi.org/10.1103/PhysRevD.65.083523} {\bibfield  {journal}
  {\bibinfo  {journal} {Phys. Rev. D}\ }\textbf {\bibinfo {volume} {65}},\
  \bibinfo {pages} {083523} (\bibinfo {year} {2002})}\BibitemShut {NoStop}%
\bibitem [{\citenamefont {Klatt}\ \emph {et~al.}(2020)\citenamefont {Klatt},
  \citenamefont {Kim},\ and\ \citenamefont {Torquato}}]{klatt_cloaking_2020}%
  \BibitemOpen
  \bibfield  {author} {\bibinfo {author} {\bibfnamefont {M.~A.}\ \bibnamefont
  {Klatt}}, \bibinfo {author} {\bibfnamefont {J.}~\bibnamefont {Kim}},\ and\
  \bibinfo {author} {\bibfnamefont {S.}~\bibnamefont {Torquato}},\ }\href
  {https://doi.org/10.1103/PhysRevE.101.032118} {\bibfield  {journal} {\bibinfo
   {journal} {Phys. Rev. E}\ }\textbf {\bibinfo {volume} {101}},\ \bibinfo
  {pages} {032118} (\bibinfo {year} {2020})}\BibitemShut {NoStop}%
\bibitem [{\citenamefont {Kubala}\ \emph {et~al.}(2022)\citenamefont {Kubala},
  \citenamefont {Batys}, \citenamefont {Barbasz}, \citenamefont
  {Wero{\'n}ski},\ and\ \citenamefont {Cie{\'s}la}}]{Kubala2022RSA}%
  \BibitemOpen
  \bibfield  {author} {\bibinfo {author} {\bibfnamefont {P.}~\bibnamefont
  {Kubala}}, \bibinfo {author} {\bibfnamefont {P.}~\bibnamefont {Batys}},
  \bibinfo {author} {\bibfnamefont {J.}~\bibnamefont {Barbasz}}, \bibinfo
  {author} {\bibfnamefont {P.}~\bibnamefont {Wero{\'n}ski}},\ and\ \bibinfo
  {author} {\bibfnamefont {M.}~\bibnamefont {Cie{\'s}la}},\ }\href
  {https://doi.org/10.1016/j.cis.2022.102692} {\bibfield  {journal} {\bibinfo
  {journal} {Adv. Colloid Interface Sci.}\ }\textbf {\bibinfo {volume} {306}},\
  \bibinfo {pages} {102692} (\bibinfo {year} {2022})}\BibitemShut {NoStop}%
\bibitem [{\citenamefont {Zhang}\ and\ \citenamefont
  {Torquato}(2013)}]{ZhangTorquato2013}%
  \BibitemOpen
  \bibfield  {author} {\bibinfo {author} {\bibfnamefont {G.}~\bibnamefont
  {Zhang}}\ and\ \bibinfo {author} {\bibfnamefont {S.}~\bibnamefont
  {Torquato}},\ }\href {https://doi.org/10.1103/PhysRevE.88.053312} {\bibfield
  {journal} {\bibinfo  {journal} {Phys. Rev. E}\ }\textbf {\bibinfo {volume}
  {88}},\ \bibinfo {pages} {053312} (\bibinfo {year} {2013})}\BibitemShut
  {NoStop}%
\bibitem [{\citenamefont {Torquato}(2026)}]{torquato2026}%
  \BibitemOpen
  \bibfield  {author} {\bibinfo {author} {\bibfnamefont {S.}~\bibnamefont
  {Torquato}},\ }\href
  {https://doi.org/https://doi.org/10.1038/s42254-026-00925-w} {\bibfield
  {journal} {\bibinfo  {journal} {Nat. Rev. Phys.}\ }\textbf {\bibinfo {volume}
  {8}},\ \bibinfo {pages} {261} (\bibinfo {year} {2026})}\BibitemShut {NoStop}%
\bibitem [{\citenamefont {Torquato}\ \emph {et~al.}(2006)\citenamefont
  {Torquato}, \citenamefont {Uche},\ and\ \citenamefont
  {Stillinger}}]{TorquatoUcheStillinger2006}%
  \BibitemOpen
  \bibfield  {author} {\bibinfo {author} {\bibfnamefont {S.}~\bibnamefont
  {Torquato}}, \bibinfo {author} {\bibfnamefont {O.~U.}\ \bibnamefont {Uche}},\
  and\ \bibinfo {author} {\bibfnamefont {F.~H.}\ \bibnamefont {Stillinger}},\
  }\href {https://doi.org/10.1103/PhysRevE.74.061308} {\bibfield  {journal}
  {\bibinfo  {journal} {Phys. Rev. E}\ }\textbf {\bibinfo {volume} {74}},\
  \bibinfo {pages} {061308} (\bibinfo {year} {2006})}\BibitemShut {NoStop}%
\bibitem [{\citenamefont {Redenbach}\ and\ \citenamefont
  {Jung}(2025)}]{Redenbach2025}%
  \BibitemOpen
  \bibfield  {author} {\bibinfo {author} {\bibfnamefont {C.}~\bibnamefont
  {Redenbach}}\ and\ \bibinfo {author} {\bibfnamefont {C.}~\bibnamefont
  {Jung}},\ }in\ \href {https://doi.org/10.1007/978-3-031-87264-8_2} {\emph
  {\bibinfo {booktitle} {Stochastic Geometry: Percolation, Tesselations,
  Gaussian Fields and Point Processes}}},\ \bibinfo {editor} {edited by\
  \bibinfo {editor} {\bibfnamefont {H.}~\bibnamefont {Bierm{\'e}}}}\ (\bibinfo
  {publisher} {Springer},\ \bibinfo {address} {Cham, Switzerland},\ \bibinfo
  {year} {2025})\ pp.\ \bibinfo {pages} {35--80}\BibitemShut {NoStop}%
\bibitem [{\citenamefont {Chen}\ \emph {et~al.}(2024)\citenamefont {Chen},
  \citenamefont {Miao}, \citenamefont {Ma}, \citenamefont {Hopkins},\ and\
  \citenamefont {Huang}}]{Chen2024MNN}%
  \BibitemOpen
  \bibfield  {author} {\bibinfo {author} {\bibfnamefont {J.}~\bibnamefont
  {Chen}}, \bibinfo {author} {\bibfnamefont {X.}~\bibnamefont {Miao}}, \bibinfo
  {author} {\bibfnamefont {H.}~\bibnamefont {Ma}}, \bibinfo {author}
  {\bibfnamefont {J.~B.}\ \bibnamefont {Hopkins}},\ and\ \bibinfo {author}
  {\bibfnamefont {G.}~\bibnamefont {Huang}},\ }\href
  {https://doi.org/10.1016/j.matdes.2024.113093} {\bibfield  {journal}
  {\bibinfo  {journal} {Materials \& Design}\ }\textbf {\bibinfo {volume}
  {244}},\ \bibinfo {pages} {113093} (\bibinfo {year} {2024})}\BibitemShut
  {NoStop}%
\bibitem [{\citenamefont {Cronau}\ and\ \citenamefont
  {Engstler}(2025)}]{Cronau2025}%
  \BibitemOpen
  \bibfield  {author} {\bibinfo {author} {\bibfnamefont {J.}~\bibnamefont
  {Cronau}}\ and\ \bibinfo {author} {\bibfnamefont {F.}~\bibnamefont
  {Engstler}},\ }\href {https://doi.org/10.1007/s40964-025-01094-5} {\bibfield
  {journal} {\bibinfo  {journal} {Prog. Addit. Manuf.}\ }\textbf {\bibinfo
  {volume} {10}},\ \bibinfo {pages} {3145} (\bibinfo {year}
  {2025})}\BibitemShut {NoStop}%
\bibitem [{\citenamefont {Du}\ \emph {et~al.}(1999)\citenamefont {Du},
  \citenamefont {Faber},\ and\ \citenamefont {Gunzburger}}]{du1999}%
  \BibitemOpen
  \bibfield  {author} {\bibinfo {author} {\bibfnamefont {Q.}~\bibnamefont
  {Du}}, \bibinfo {author} {\bibfnamefont {V.}~\bibnamefont {Faber}},\ and\
  \bibinfo {author} {\bibfnamefont {M.}~\bibnamefont {Gunzburger}},\ }\href
  {https://doi.org/10.1137/S0036144599352836} {\bibfield  {journal} {\bibinfo
  {journal} {SIAM Rev.}\ }\textbf {\bibinfo {volume} {41}},\ \bibinfo {pages}
  {637} (\bibinfo {year} {1999})}\BibitemShut {NoStop}%
\bibitem [{\citenamefont {do~Carmo}(1976)}]{doCarmo1976}%
  \BibitemOpen
  \bibfield  {author} {\bibinfo {author} {\bibfnamefont {M.~P.}\ \bibnamefont
  {do~Carmo}},\ }\href@noop {} {\emph {\bibinfo {title} {Differential Geometry
  of Curves and Surfaces}}}\ (\bibinfo  {publisher} {Prentice-Hall},\ \bibinfo
  {address} {Sacramento, CA, USA},\ \bibinfo {year} {1976})\BibitemShut
  {NoStop}%
\bibitem [{\citenamefont {Ollivier}(2007)}]{ollivier2007}%
  \BibitemOpen
  \bibfield  {author} {\bibinfo {author} {\bibfnamefont {Y.}~\bibnamefont
  {Ollivier}},\ }\href
  {https://doi.org/https://doi.org/10.1016/j.crma.2007.10.041} {\bibfield
  {journal} {\bibinfo  {journal} {C. R. Acad. Sci. Paris, Ser. I}\ }\textbf
  {\bibinfo {volume} {345}},\ \bibinfo {pages} {643} (\bibinfo {year}
  {2007})}\BibitemShut {NoStop}%
\bibitem [{\citenamefont {Lin}\ \emph {et~al.}(2011)\citenamefont {Lin},
  \citenamefont {Lu},\ and\ \citenamefont {Yau}}]{lin2011}%
  \BibitemOpen
  \bibfield  {author} {\bibinfo {author} {\bibfnamefont {Y.}~\bibnamefont
  {Lin}}, \bibinfo {author} {\bibfnamefont {L.}~\bibnamefont {Lu}},\ and\
  \bibinfo {author} {\bibfnamefont {S.~T.}\ \bibnamefont {Yau}},\ }\href
  {https://doi.org/https://doi.org/10.2748/tmj/1325886283} {\bibfield
  {journal} {\bibinfo  {journal} {Tohoku Math. J.}\ }\textbf {\bibinfo {volume}
  {63}},\ \bibinfo {pages} {605} (\bibinfo {year} {2011})}\BibitemShut
  {NoStop}%
\bibitem [{\citenamefont {Chandra}\ \emph {et~al.}(1996)\citenamefont
  {Chandra}, \citenamefont {Raghavan}, \citenamefont {Ruzzo}, \citenamefont
  {Smolensky},\ and\ \citenamefont {Tiwari}}]{ChandraEtAl1989}%
  \BibitemOpen
  \bibfield  {author} {\bibinfo {author} {\bibfnamefont {A.~K.}\ \bibnamefont
  {Chandra}}, \bibinfo {author} {\bibfnamefont {P.}~\bibnamefont {Raghavan}},
  \bibinfo {author} {\bibfnamefont {W.~L.}\ \bibnamefont {Ruzzo}}, \bibinfo
  {author} {\bibfnamefont {R.}~\bibnamefont {Smolensky}},\ and\ \bibinfo
  {author} {\bibfnamefont {P.}~\bibnamefont {Tiwari}},\ }\href
  {https://doi.org/https://doi.org/10.1007/BF01270385} {\bibfield  {journal}
  {\bibinfo  {journal} {Comput. Complex.}\ }\textbf {\bibinfo {volume} {6}},\
  \bibinfo {pages} {312} (\bibinfo {year} {1996})}\BibitemShut {NoStop}%
\bibitem [{\citenamefont {Wang}\ \emph {et~al.}(2014)\citenamefont {Wang},
  \citenamefont {Pournaras}, \citenamefont {Kooij},\ and\ \citenamefont
  {{Van~Mieghem}}}]{Wang2014}%
  \BibitemOpen
  \bibfield  {author} {\bibinfo {author} {\bibfnamefont {X.}~\bibnamefont
  {Wang}}, \bibinfo {author} {\bibfnamefont {E.}~\bibnamefont {Pournaras}},
  \bibinfo {author} {\bibfnamefont {R.~E.}\ \bibnamefont {Kooij}},\ and\
  \bibinfo {author} {\bibfnamefont {P.}~\bibnamefont {{Van~Mieghem}}},\ }\href
  {https://doi.org/10.1140/epjb/e2014-50276-0} {\bibfield  {journal} {\bibinfo
  {journal} {Eur. Phys. J. B}\ }\textbf {\bibinfo {volume} {87}},\ \bibinfo
  {pages} {221} (\bibinfo {year} {2014})}\BibitemShut {NoStop}%
\bibitem [{\citenamefont {Ghosh}\ \emph {et~al.}(2008)\citenamefont {Ghosh},
  \citenamefont {Boyd},\ and\ \citenamefont {Saberi}}]{Ghosh2008}%
  \BibitemOpen
  \bibfield  {author} {\bibinfo {author} {\bibfnamefont {A.}~\bibnamefont
  {Ghosh}}, \bibinfo {author} {\bibfnamefont {S.}~\bibnamefont {Boyd}},\ and\
  \bibinfo {author} {\bibfnamefont {A.}~\bibnamefont {Saberi}},\ }\href
  {https://doi.org/10.1137/060651492} {\bibfield  {journal} {\bibinfo
  {journal} {SIAM Review}\ }\textbf {\bibinfo {volume} {50}},\ \bibinfo {pages}
  {37} (\bibinfo {year} {2008})}\BibitemShut {NoStop}%
\bibitem [{\citenamefont {Masuda}\ \emph {et~al.}(2017)\citenamefont {Masuda},
  \citenamefont {Porter},\ and\ \citenamefont
  {Lambiotte}}]{MasudaPorterLambiotte2017}%
  \BibitemOpen
  \bibfield  {author} {\bibinfo {author} {\bibfnamefont {N.}~\bibnamefont
  {Masuda}}, \bibinfo {author} {\bibfnamefont {M.~A.}\ \bibnamefont {Porter}},\
  and\ \bibinfo {author} {\bibfnamefont {R.}~\bibnamefont {Lambiotte}},\ }\href
  {https://doi.org/https://doi.org/10.1016/j.physrep.2017.07.007} {\bibfield
  {journal} {\bibinfo  {journal} {Physics Reports}\ }\textbf {\bibinfo {volume}
  {716--717}},\ \bibinfo {pages} {1} (\bibinfo {year} {2017})}\BibitemShut
  {NoStop}%
\bibitem [{\citenamefont {Kelathaya}\ \emph {et~al.}(2023)\citenamefont
  {Kelathaya}, \citenamefont {Bapat},\ and\ \citenamefont
  {Karantha}}]{KelathayaBapatKarantha2023}%
  \BibitemOpen
  \bibfield  {author} {\bibinfo {author} {\bibfnamefont {U.}~\bibnamefont
  {Kelathaya}}, \bibinfo {author} {\bibfnamefont {R.~B.}\ \bibnamefont
  {Bapat}},\ and\ \bibinfo {author} {\bibfnamefont {M.~P.}\ \bibnamefont
  {Karantha}},\ }\href {https://doi.org/10.1080/09728600.2023.2234002}
  {\bibfield  {journal} {\bibinfo  {journal} {AKCE Int. J. Graphs Combin.}\
  }\textbf {\bibinfo {volume} {20}},\ \bibinfo {pages} {108} (\bibinfo {year}
  {2023})}\BibitemShut {NoStop}%
\bibitem [{\citenamefont {Levin}\ and\ \citenamefont
  {Peres}(2017)}]{LevinPeresWilmer2017}%
  \BibitemOpen
  \bibfield  {author} {\bibinfo {author} {\bibfnamefont {D.~A.}\ \bibnamefont
  {Levin}}\ and\ \bibinfo {author} {\bibfnamefont {Y.}~\bibnamefont {Peres}},\
  }\href@noop {} {\emph {\bibinfo {title} {Markov Chains and Mixing Times}}},\
  \bibinfo {edition} {2nd}\ ed.\ (\bibinfo  {publisher} {American Mathematical
  Society},\ \bibinfo {address} {Providence, RI, USA},\ \bibinfo {year}
  {2017})\ \bibinfo {note} {\hspace{-1.5mm}; with contributions by E.~L.
  Wilmer}\BibitemShut {NoStop}%
\bibitem [{\citenamefont {Schwarze}\ \emph {et~al.}(2024)\citenamefont
  {Schwarze}, \citenamefont {Jiang}, \citenamefont {Wray},\ and\ \citenamefont
  {Porter}}]{Schwarze2024}%
  \BibitemOpen
  \bibfield  {author} {\bibinfo {author} {\bibfnamefont {A.~C.}\ \bibnamefont
  {Schwarze}}, \bibinfo {author} {\bibfnamefont {J.}~\bibnamefont {Jiang}},
  \bibinfo {author} {\bibfnamefont {J.}~\bibnamefont {Wray}},\ and\ \bibinfo
  {author} {\bibfnamefont {M.~A.}\ \bibnamefont {Porter}},\ }\href@noop {}
  {\bibfield  {journal} {\bibinfo  {journal} {arXiv preprint}\ } (\bibinfo
  {year} {2024})},\ \bibinfo {note} {arXiv:2409.07498}\BibitemShut {NoStop}%
\bibitem [{\citenamefont {Artime}\ \emph {et~al.}(2024)\citenamefont {Artime},
  \citenamefont {Grassia}, \citenamefont {Domenico}, \citenamefont {Gleeson},
  \citenamefont {Makse}, \citenamefont {Mangioni}, \citenamefont {Perc},\ and\
  \citenamefont {Radicchi}}]{Artime2024}%
  \BibitemOpen
  \bibfield  {author} {\bibinfo {author} {\bibfnamefont {O.}~\bibnamefont
  {Artime}}, \bibinfo {author} {\bibfnamefont {M.}~\bibnamefont {Grassia}},
  \bibinfo {author} {\bibfnamefont {M.~D.}\ \bibnamefont {Domenico}}, \bibinfo
  {author} {\bibfnamefont {J.~P.}\ \bibnamefont {Gleeson}}, \bibinfo {author}
  {\bibfnamefont {H.~A.}\ \bibnamefont {Makse}}, \bibinfo {author}
  {\bibfnamefont {G.}~\bibnamefont {Mangioni}}, \bibinfo {author}
  {\bibfnamefont {M.}~\bibnamefont {Perc}},\ and\ \bibinfo {author}
  {\bibfnamefont {F.}~\bibnamefont {Radicchi}},\ }\href
  {https://doi.org/10.1038/s42254-023-00676-y} {\bibfield  {journal} {\bibinfo
  {journal} {Nat. Rev. Phys.}\ }\textbf {\bibinfo {volume} {6}},\ \bibinfo
  {pages} {114} (\bibinfo {year} {2024})}\BibitemShut {NoStop}%
\bibitem [{\citenamefont {Shi}\ \emph {et~al.}(2023)\citenamefont {Shi},
  \citenamefont {Keeney}, \citenamefont {Chen}, \citenamefont {Jiao},\ and\
  \citenamefont {Torquato}}]{shi_keeney_chen_jiao_torquato_2023}%
  \BibitemOpen
  \bibfield  {author} {\bibinfo {author} {\bibfnamefont {W.}~\bibnamefont
  {Shi}}, \bibinfo {author} {\bibfnamefont {D.}~\bibnamefont {Keeney}},
  \bibinfo {author} {\bibfnamefont {D.}~\bibnamefont {Chen}}, \bibinfo {author}
  {\bibfnamefont {Y.}~\bibnamefont {Jiao}},\ and\ \bibinfo {author}
  {\bibfnamefont {S.}~\bibnamefont {Torquato}},\ }\href
  {https://doi.org/10.1103/PhysRevE.108.045306} {\bibfield  {journal} {\bibinfo
   {journal} {Phys. Rev. E}\ }\textbf {\bibinfo {volume} {108}},\ \bibinfo
  {pages} {045306} (\bibinfo {year} {2023})}\BibitemShut {NoStop}%
\bibitem [{\citenamefont {Chen}\ \emph {et~al.}(2018)\citenamefont {Chen},
  \citenamefont {Lomba},\ and\ \citenamefont
  {Torquato}}]{chen_lomba_torquato_2018}%
  \BibitemOpen
  \bibfield  {author} {\bibinfo {author} {\bibfnamefont {D.}~\bibnamefont
  {Chen}}, \bibinfo {author} {\bibfnamefont {E.}~\bibnamefont {Lomba}},\ and\
  \bibinfo {author} {\bibfnamefont {S.}~\bibnamefont {Torquato}},\ }\href
  {https://doi.org/10.1039/C8CP02616E} {\bibfield  {journal} {\bibinfo
  {journal} {Phys. Chem. Chem. Phys.}\ }\textbf {\bibinfo {volume} {20}},\
  \bibinfo {pages} {17557} (\bibinfo {year} {2018})}\BibitemShut {NoStop}%
\bibitem [{\citenamefont {Ecevit}\ and\ \citenamefont
  {Boysal}(2025)}]{EcevitBoysal2025}%
  \BibitemOpen
  \bibfield  {author} {\bibinfo {author} {\bibfnamefont {F.}~\bibnamefont
  {Ecevit}}\ and\ \bibinfo {author} {\bibfnamefont {A.}~\bibnamefont
  {Boysal}},\ }\href {https://doi.org/10.55730/1300-0098.3571} {\bibfield
  {journal} {\bibinfo  {journal} {Turk. J. Math.}\ }\textbf {\bibinfo {volume}
  {49}},\ \bibinfo {pages} {18} (\bibinfo {year} {2025})}\BibitemShut {NoStop}%
\bibitem [{\citenamefont {Ye}(2011)}]{Ye01062011}%
  \BibitemOpen
  \bibfield  {author} {\bibinfo {author} {\bibfnamefont {L.}~\bibnamefont
  {Ye}},\ }\href {https://doi.org/10.1080/03081081003794233} {\bibfield
  {journal} {\bibinfo  {journal} {Lin. Multilin. Alg.}\ }\textbf {\bibinfo
  {volume} {59}},\ \bibinfo {pages} {645} (\bibinfo {year} {2011})}\BibitemShut
  {NoStop}%
\bibitem [{\citenamefont {Moody}\ \emph
  {et~al.}(2026{\natexlab{b}})\citenamefont {Moody}, \citenamefont {Li},
  \citenamefont {Maher}, \citenamefont {Lee}, \citenamefont {Horn},
  \citenamefont {Newhall}, \citenamefont {Hurley}, \citenamefont {Daniels},\
  and\ \citenamefont {Rock}}]{rock2025}%
  \BibitemOpen
  \bibfield  {author} {\bibinfo {author} {\bibfnamefont {K.}~\bibnamefont
  {Moody}}, \bibinfo {author} {\bibfnamefont {M.}~\bibnamefont {Li}}, \bibinfo
  {author} {\bibfnamefont {C.~E.}\ \bibnamefont {Maher}}, \bibinfo {author}
  {\bibfnamefont {K.}~\bibnamefont {Lee}}, \bibinfo {author} {\bibfnamefont
  {T.}~\bibnamefont {Horn}}, \bibinfo {author} {\bibfnamefont {K.~A.}\
  \bibnamefont {Newhall}}, \bibinfo {author} {\bibfnamefont {R.}~\bibnamefont
  {Hurley}}, \bibinfo {author} {\bibfnamefont {K.~E.}\ \bibnamefont
  {Daniels}},\ and\ \bibinfo {author} {\bibfnamefont {C.}~\bibnamefont {Rock}}}
  (\bibinfo {year} {2026}{\natexlab{b}}),\ \bibinfo {note} {engrXiv preprint;
  available at \url{https://engrxiv.org/preprint/view/5865}}\BibitemShut
  {NoStop}%
\bibitem [{\citenamefont {Ni}\ \emph {et~al.}(2015)\citenamefont {Ni},
  \citenamefont {Lin}, \citenamefont {Gao}, \citenamefont {Gu},\ and\
  \citenamefont {Saucan}}]{Ni2015}%
  \BibitemOpen
  \bibfield  {author} {\bibinfo {author} {\bibfnamefont {C.-C.}\ \bibnamefont
  {Ni}}, \bibinfo {author} {\bibfnamefont {Y.-Y.}\ \bibnamefont {Lin}},
  \bibinfo {author} {\bibfnamefont {J.}~\bibnamefont {Gao}}, \bibinfo {author}
  {\bibfnamefont {X.~D.}\ \bibnamefont {Gu}},\ and\ \bibinfo {author}
  {\bibfnamefont {E.}~\bibnamefont {Saucan}},\ }in\ \href
  {https://doi.org/10.1109/INFOCOM.2015.7218668} {\emph {\bibinfo {booktitle}
  {Proc. IEEE Conf. Comp. Comm. (INFOCOM) 2015}}}\ (\bibinfo {year} {2015})\
  pp.\ \bibinfo {pages} {2758--2766}\BibitemShut {NoStop}%
\bibitem [{\citenamefont {Sandhu}\ \emph {et~al.}(2016)\citenamefont {Sandhu},
  \citenamefont {Georgiou},\ and\ \citenamefont {Tannenbaum}}]{Sandhu2016}%
  \BibitemOpen
  \bibfield  {author} {\bibinfo {author} {\bibfnamefont {R.~S.}\ \bibnamefont
  {Sandhu}}, \bibinfo {author} {\bibfnamefont {T.~T.}\ \bibnamefont
  {Georgiou}},\ and\ \bibinfo {author} {\bibfnamefont {A.~R.}\ \bibnamefont
  {Tannenbaum}},\ }\href {https://doi.org/10.1126/sciadv.1501495} {\bibfield
  {journal} {\bibinfo  {journal} {Sci. Adv.}\ }\textbf {\bibinfo {volume}
  {2}},\ \bibinfo {pages} {e1501495} (\bibinfo {year} {2016})}\BibitemShut
  {NoStop}%
\bibitem [{\citenamefont {Gosztolai}\ and\ \citenamefont
  {Arnaudon}(2021)}]{Gosztolai2021}%
  \BibitemOpen
  \bibfield  {author} {\bibinfo {author} {\bibfnamefont {A.}~\bibnamefont
  {Gosztolai}}\ and\ \bibinfo {author} {\bibfnamefont {A.}~\bibnamefont
  {Arnaudon}},\ }\href {https://doi.org/10.1038/s41467-021-24884-1} {\bibfield
  {journal} {\bibinfo  {journal} {Nat. Comm.}\ }\textbf {\bibinfo {volume}
  {12}},\ \bibinfo {pages} {4561} (\bibinfo {year} {2021})}\BibitemShut
  {NoStop}%
\bibitem [{\citenamefont {Lee}\ \emph {et~al.}(2014)\citenamefont {Lee},
  \citenamefont {Cucuringu},\ and\ \citenamefont {Porter}}]{lee2014}%
  \BibitemOpen
  \bibfield  {author} {\bibinfo {author} {\bibfnamefont {S.~H.}\ \bibnamefont
  {Lee}}, \bibinfo {author} {\bibfnamefont {M.}~\bibnamefont {Cucuringu}},\
  and\ \bibinfo {author} {\bibfnamefont {M.~A.}\ \bibnamefont {Porter}},\
  }\href {https://doi.org/10.1103/PhysRevE.89.032810} {\bibfield  {journal}
  {\bibinfo  {journal} {Phys. Rev. E}\ }\textbf {\bibinfo {volume} {89}},\
  \bibinfo {pages} {032810} (\bibinfo {year} {2014})}\BibitemShut {NoStop}%
\bibitem [{\citenamefont {Cetinay}\ \emph {et~al.}(2018)\citenamefont
  {Cetinay}, \citenamefont {Devriendt},\ and\ \citenamefont
  {Mieghem}}]{dev2018}%
  \BibitemOpen
  \bibfield  {author} {\bibinfo {author} {\bibfnamefont {H.}~\bibnamefont
  {Cetinay}}, \bibinfo {author} {\bibfnamefont {K.}~\bibnamefont {Devriendt}},\
  and\ \bibinfo {author} {\bibfnamefont {P.~V.}\ \bibnamefont {Mieghem}},\
  }\href {https://doi.org/https://doi.org/10.1007/s41109-018-0089-9} {\bibfield
   {journal} {\bibinfo  {journal} {Appl. Netw. Sci.}\ }\textbf {\bibinfo
  {volume} {3}},\ \bibinfo {pages} {34} (\bibinfo {year} {2018})}\BibitemShut
  {NoStop}%
\bibitem [{\citenamefont {Yang}\ \emph {et~al.}(2025)\citenamefont {Yang},
  \citenamefont {Yi}, \citenamefont {Li}, \citenamefont {Li}, \citenamefont
  {Ye}, \citenamefont {Li},\ and\ \citenamefont
  {Christensen}}]{Yang2025_cloaking}%
  \BibitemOpen
  \bibfield  {author} {\bibinfo {author} {\bibfnamefont {Z.}~\bibnamefont
  {Yang}}, \bibinfo {author} {\bibfnamefont {J.}~\bibnamefont {Yi}}, \bibinfo
  {author} {\bibfnamefont {F.}~\bibnamefont {Li}}, \bibinfo {author}
  {\bibfnamefont {Z.}~\bibnamefont {Li}}, \bibinfo {author} {\bibfnamefont
  {L.}~\bibnamefont {Ye}}, \bibinfo {author} {\bibfnamefont {B.}~\bibnamefont
  {Li}},\ and\ \bibinfo {author} {\bibfnamefont {J.}~\bibnamefont
  {Christensen}},\ }\href {https://doi.org/10.1038/s41467-025-63939-5}
  {\bibfield  {journal} {\bibinfo  {journal} {Nat. Commun.}\ }\textbf {\bibinfo
  {volume} {16}},\ \bibinfo {pages} {8858} (\bibinfo {year}
  {2025})}\BibitemShut {NoStop}%
\end{thebibliography}

%%%%


\end{document}